\documentclass[aps,10pt,twocolumn,superscriptaddress,footinbib,flushbottom]{revtex4-1}%
\usepackage{amssymb,amsmath,amsfonts,bm,dcolumn}
\usepackage{graphicx,color}
\usepackage[squaren]{SIunits}
\usepackage[english]{babel}
\usepackage{tabularx}
\usepackage{tikz}

\renewcommand{\vec}[1]{\boldsymbol{\mathrm{#1}}}%
\newcommand{\uvec}[1]{\bm{\hat{\mathbf{#1}}}}
\newcommand{\abs}[1]{\lvert#1\rvert}%
\newcommand{\smallfrac}[2]{\textstyle\frac{#1}{#2}}%

\newcommand{\mean}[1]{\langle #1 \rangle}

\renewcommand{\O}[1]{\mathcal{O}\left(#1\right)}

\newcommand\id{\bm{\ensuremath{\mathbb{I}}}}

\newcommand{\dif}{d}

\newcommand{\tdif}[2]{\frac{\dif  #1}{\dif  #2}}%

\newcommand{\integral}[3]{\int_{#1}^{#2}\hspace{-0.05cm}\dif{#3}} 

\renewcommand{\Re}{\fct{Re}}%

\newcommand{\fct}[1]{\operatorname{#1}}

\newmuskip\pFqmuskip
\newcommand*\pFq[6][8]{%
	\begingroup 
	\pFqmuskip=#1mu\relax
	\mathchardef\normalcomma=\mathcode`,
	\mathcode`\,=\string"8000
	\begingroup\lccode`\~=`\,
	\lowercase{\endgroup\let~}\pFqcomma
	{}_{#2}F_{#3}{\left[\genfrac..{0pt}{}{#4}{#5};#6\right]}%
	\endgroup
}
\newcommand{\pFqcomma}{{\normalcomma}\mskip\pFqmuskip}

\begin{document}

\title{Time-dependent inertia of self-propelled particles: the Langevin rocket}

\author{Alexander R. Sprenger}
\thanks{These two authors contributed equally}
\affiliation{Institut f\"{u}r Theoretische Physik II: Weiche Materie, Heinrich-Heine-Universit\"{a}t D\"{u}sseldorf, D-40225 D\"{u}sseldorf, Germany}

\author{Soudeh Jahanshahi}
\thanks{These two authors contributed equally}
\affiliation{Institut f\"{u}r Theoretische Physik II: Weiche Materie, Heinrich-Heine-Universit\"{a}t D\"{u}sseldorf, D-40225 D\"{u}sseldorf, Germany}

\author{Alexei V. Ivlev}
\affiliation{Max-Planck-Institut f{\"u}r Extraterrestrische Physik, 85748 Garching, Germany}

\author{Hartmut L\"{o}wen}
\affiliation{Institut f\"{u}r Theoretische Physik II: Weiche Materie, Heinrich-Heine-Universit\"{a}t D\"{u}sseldorf, D-40225 D\"{u}sseldorf, Germany}


\begin{abstract}
	Many self-propelled objects are large enough to exhibit inertial effects but still suffer from environmental fluctuations. The corresponding basic equations of motion are governed by active Langevin dynamics which involve inertia, friction and stochastic noise for both the translational and orientational degrees of freedom coupled via the self-propulsion along the particle orientation.
	In this paper, we generalize the active Langevin model to time-dependent parameters and explicitly discuss the effect of time-dependent inertia for achiral and chiral particles. 
	Realizations of this situation are manifold ranging from minirockets which are self-propelled by burning their own mass, dust particles in plasma which lose mass by evaporating material to walkers with expiring activity. 
	Here we present analytical solutions for several dynamical correlation functions such as the mean-square displacement and the orientational and velocity autocorrelation functions.
	If the parameters exhibit a slow power-law in time, we obtain anomalous superdiffusion with a non-trivial dynamical exponent.
	Finally we constitute the ``Langevin rocket" model by including  orientational fluctuations in the  traditional Tsiolkovsky rocket equation. We calculate
	the mean reach of the Langevin rocket 
	and discuss different mass ejection strategies to maximize it.
	Our results can be tested in experiments on  macroscopic robotic or living particles or in self-propelled mesoscopic objects  
	moving in media of low viscosity such as complex plasma.
\end{abstract}

\maketitle

\section{Introduction}

The non-equilibrium dynamics of active Brownian particles --also referred to as microswimmers-- are typically described in the overdamped limit, where inertial effects are sufficiently small relative to viscous ones \cite{HowseJRGVG2007,RomanczukBELSG2012,ElgetiWG2015,BechingerDLLRVV2016}. 
This is an excellent approximation for micron-sized self-propelled particles swimming in a viscous Newtonian liquid such as water \cite{Purcell1977} at low Reynolds number. 
The standard model of a single active Brownian particle \cite{HowseJRGVG2007,tenHagenTL2011,BechingerDLLRVV2016} involves a translational and an orientational degree of freedom and includes Stokesian friction and fluctuations.
These degrees of freedom are coupled via the self-propulsion along the particle orientation which is modeled in a simple averaged way by an internal velocity, sometimes referred to as the particle activity.

However, inertial effects become relevant for larger particle sizes or the motion in gaseous media of lower viscosity. 
Though highly relevant for swimming and flying organisms  as well as for autonomous machines (e.g., flying insect-drones, marine robots) \cite{Klotsa2019} mesoscale active matter at intermediate Reynolds number has been much less studied. 
Aiming at a simple description of a single particle first, one basic model is that of {\it active Langevin motion} \cite{ScholzJLL2018,CallegariV2019,UmSJ2019,Loewen2020}: it generalizes the common overdamped model of active Brownian motion \cite{HowseJRGVG2007,tenHagenTL2011,BechingerDLLRVV2016} towards underdamped dynamics by including finite particle mass and moment of inertia in the equations of motion \cite{EnculescuS2011,GhoshLMM2015,JoyeuxB2016,ManacordaP2017,TakatoriB2017,MokhtariAZ2017,DasGW2019,Sandoval2020,GutierrezS2020}.
The inertial self-propelled particles may therefore be called ``microflyers'' (rather than ``microswimmers''), sometimes they are also termed ``runners'', ``walkers''  or ``hoppers'' \cite{EbelingST1999,WeberHDLDFC2013,ParisiCHZ2018}.
Examples for inanimate inertial self-propelled particles modeled by active Langevin dynamics are manifold, they include a complex plasma consisting of mesoscopic dust particles in a weakly ionized gas  \cite{MorfillI2009,SuetterlinEtAl2009,CoueedelNIZTM2010,ChaudhuriIKTM2011,IvlevBHDNL2015,NosenkoLKRZT2020}, vibration-driven granular particles \cite{NarayanRM2007,KudrolliLVT2008,DeseigneDC2010,GiomiHWM2013,WeberHDLDFC2013,KlotsaBHBS2015,PattersonFSJKGZPP2017,JunotBLD2017,Ramaswamy2017,DeblaisBGDVLBBK2018,DauchotD2019}, autorotating seeds and fruits \cite{RabaultFC2019,FauliRC2019}, camphor surfers \cite{LeoniPEENASA2020}, hexbug crawlers\cite{LeoniPEENASA2020}, trapped aerosols \cite{DiLeonardoRLPWGBM2007} and mini-robots \cite{RubensteinCN2014,FujiwaraKI2014,TolbaAR2015,ZhakypovMHP2019,YangRCZ2019}.
Moreover, there are numerous examples of animals  moving at intermediate Reynolds number such as swimming organisms (nematodes, brine shrimps, whirligig beetles etc) \cite{Klotsa2019,Turner2019} and  flying insects and birds \cite{TonerTu1995,Flocks1998,Chiappini2008,Bartussek2016,Mukundarajan2016,Bartussek2018,Attanasi2014}.

In this paper, we extend the active Langevin model to {\it time-dependent} parameters such as time-dependent inertia, self-propulsion and friction. 
This is a situation frequently encountered in nature and realizable in laboratory experiments on artificial self-propelled objects. 
Let us mention some examples: scallops move their shells and accelerate by jet propulsion. 
Therefore, they become smaller in the course of the motion such that their moment of inertia  and their friction coefficients become time-dependent.
Moving animals typically have a finite energy reservoir \cite{EbelingST1999} implying that their self-propulsion velocity is getting slower as a function of time.
The maneuverability of animal motion is provided by changes in the body shape \cite{VANDENBERG1995,Qiao2014}, which implies a change in the moment of inertia at fixed total mass.
Likewise, in the inanimate world, {\it minirockets}, which are propelled by ejecting mass, are getting lighter as a function of time \cite{KrasheninnikovPSS2010,NosenkoIM2010}.
Similarly, inflated toy balloons \cite{Mueller,ManganD2015} are self-propelled by jet propulsion and strongly subject to random fluctuations in their orientation, their body size shrinks as a function of time and so does the mass, the moment of inertia, the friction coefficient and the self-propulsion speed.
Granulate hoppers equipped with an internal vibration motor ("hexbugs") \cite{DauchotD2019}, will consume energy such that the self-propulsion speed will slowly expire and fade away as a function of time.
Robots that pick up or release objects possess a mass variation  \cite{variableMass_2002}, too, and a time-dependent mass can bring about time-dependent friction coefficients \cite{PANCHAL20081304}.
Last but not least, any prescribed time-dependence can be programmed artificially at wish for man-made robots, artificial walkers and microswimmer: the self-propulsion speed  can be made time-dependent by exposing particles to external optical fields \cite{LozanoLiebchen2019}, the noise strength can be steered by external fields \cite{RodriguezGARBVI2020,SprengerFRAIWL2020}, the damping  by the solvent viscosity \cite{PRE.rotationalViscosity2000,PRL.rotationalViscosity2018} or both by the external vibration amplitude and frequency  \cite{droplet1,KlotsaBHBS2015,droplet2,ScholzJLL2018}.

For the active Langevin model with prescribed time-dependent parameters, we present here analytical solutions for several dynamical correlation functions such as the orientational and velocity autocorrelation function, the mean displacements, the mean-square displacement and the delay function. 
Our results are as follows: First, we constitute a model which we refer to as the {\it Langevin rocket}. 
In doing so, we combine orientational fluctuations and mass loss described  by the  traditional Tsiolkovsky rocket equation \cite{Tsiolkovsky1903}.
We calculate the mean reach of the Langevin rocket 
and discuss different mass ejection strategies to maximize it. For increasing rotational noise, the optimal strategy to achieve a maximal 
reach changes discontinuously from a complete mass ejection extended over a long time to a instantaneous ejection of a mass fraction.
Second, we compare different set-ups of time-dependent inertia such as directed and isotropic mass ejection and isotropic shape changes with constant mass.
Last, we study the case of slow ("adiabatic") variation of system parameters.
In particular, for a change in the system parameters described by a power law in time we predict a superdiffusive {\it{anomalous}}  diffusion involving a mean-square displacement  $\propto t^\alpha$ which scales as a power law in time $t$ with a non-trivial exponent $\alpha$ \cite{MetzlerPRL2000,AnDiff_Gol_2009,AnDiff2013,AnDiff2014,AnDiff.PRE.2017,AnDiff2017,charalambous2019control,Chaki2019,BodrovaCCSSM2016}. In particular, we discuss chiral particles exposed to a torque which exhibit circling motion.
This generalizes earlier work for overdamped systems \cite{vanTeeffelen2008,Babel_2014,Olsen2020}.
Our predictions can be tested in various experimental set-ups ranging from macroscopic vibrated granular matter, robots or living systems to  self-propelled micron-particles that are flying in a gaseous medium or in a plasma.

The paper is organized as follows: in Sec.~\ref{sec:theory}, we introduce the  theoretical model for active Langevin motion describing an inertial particle. 
In Section~\ref{sec:constant_parameters} we recapitulate the case of  time-independent self-propulsion, inertia, damping and fluctuations found earlier \cite{GhoshLMM2015,ScholzJLL2018} but include here also new results such as an analytical expression for the time-resolved mean trajectory and mean-square-displacement.
In Sec.~\ref{sec:time_dependent_inertia}, we demonstrate how time-dependent parameters change the dynamics of the system: in particular, we introduce the Langevin rocket model and study slow temporal variations. 
Finally, we conclude in Sec.~\ref{sec:conclusion}.

\section{\label{sec:theory}Basic model and different set-ups}

\begin{figure}[t]
	\centering
	\includegraphics[width = \columnwidth]{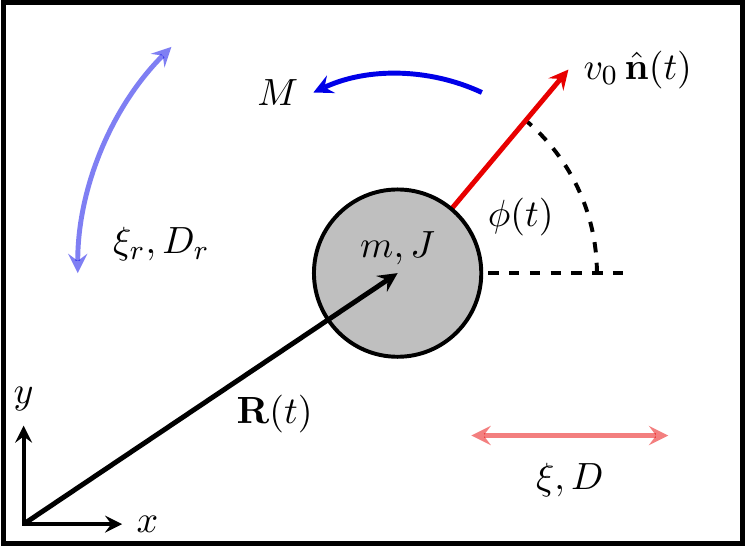}
	\caption{\label{fig:sketch}
		Self-propelled inertial particle with center position $\vec{R}(t)$
		at time $t$ moving with its center in the two-dimensional $xy$-plane. The particle position is indicated as  $\vec{R}(t)$ (black arrow).
		Moreover the particle possesses an orientational degree of freedom which is characterized by a unit vector $\uvec{n} =(\cos \phi, \sin \phi)$
		with $\phi$ denoting the angle relative to the $x$-axis. The particle self-propels along its orientation with the  velocity
		$v_0 \uvec{n}$ (red arrow). It may also experience a  torque $M$ along the $z$-axis leading to rotational motion as indicated by the blue arrow.
		The translational motion is further influenced by a  translational friction $\xi$ and the noise strength $D$ (as indicated by the light red horizontal double arrow)
		while the rotational motion is  influenced by a  rotational friction $\xi_r$ and the orientational noise strength $D_r$ (as indicated by the light blue curved double arrow). }
\end{figure}

\begin{figure}[t]
	\centering
	\includegraphics[width = \columnwidth]{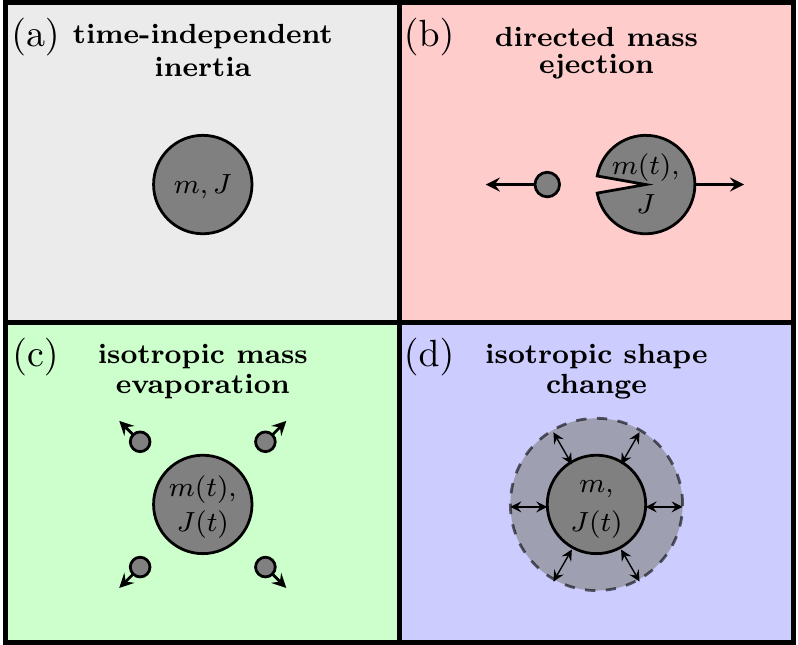}
	\caption{\label{fig:set_ups}
		Schematic illustration of the different special set-ups for an active inertial particle.
		The particle is shown as a dark-gray sphere  and its inertia
		is characterized by the particle mass $m$ and its moment of inertia $J$. (a) Time-independent inertia with constant $m$ and $J$ as a reference situation (gray background).
		(b) Directed mass ejection: Per unit time, the mass $\dot{m}(t)$ is ejected centrally with a velocity $-u \uvec{n}(t)$ along the particle orientation $\uvec{n}(t)$ which leads to a change $-u \uvec{n} \dot{m}(t)$ in the translational momentum of the particle and a time-dependent particle mass $m(t)$ but a constant moment of inertia $J$ (red background).
		(c) Isotropic mass evaporation. Here the translational and the angular momentum of the particle are both
		conserved but the particle mass $m(t)$ and moment of inertia $J(t)$ are time-dependent (green background).
		(d) Isotropic change in the particle shape.
		Here again the linear and the angular momentum of the particle are both
		conserved, the particle mass $m$ is constant but the moment of inertia $J(t)$ is time-dependent (blue background). }
\end{figure}

In this section, we define the basic model of underdamped Langevin motion for a self-propelled particle with time-dependent inertia.
We consider a self-propelled inertial particle with a center-of-mass coordinate  $\vec{R}(t)$
at time $t$ moving with its center in the two-dimensional $xy$-plane, see Figure~\ref{fig:sketch} for a sketch.
The particle is polar such that it possesses an orientational degree of freedom characterized by a unit vector $\uvec{n}(t) =(\cos \phi(t), \sin \phi(t))$
where $\phi(t)$ is the angle relative to the $x$-axis. The particle self-propels along its orientation with the self-propulsion velocity
$v_0 \uvec{n}$, also indicated in Figure~\ref{fig:sketch}. It may also  additionally be exposed to an external or internal torque $M$ along the $z$-axis  leading to
an angular velocity as shown by the blue arrow in Figure~\ref{fig:sketch}.
As the particle has inertia in both translation and rotation, its configuration fully specified by its center-of-mass coordinate  $\vec{R}(t)$, its center-of-mass velocity $\dot{\vec{R}}(t) = \dif  \vec{R}(t)/ \dif t$, its orientational angle $\phi(t)$ and its angular velocity $\dot{\phi}(t)$. 

While previous work \cite{ScholzJLL2018,Loewen2020,Sandoval2020} has considered constant particle mass and moment of inertia, here we generalize the model towards time-dependent parameters with a particular focus on a time-dependent particle mass $m(t)$ and a time-dependent moment of inertia $J(t)$, which we define with respect to the center-of-mass to describe the rotation around the $z$-axis. 
It turns out that the corresponding equations of motion need to be discussed with care as they depend on the physical origin of the
change in inertia. In order to do this systematically step by step, we first consider four different set-ups which are outlined in Figure~\ref{fig:set_ups} and which are actually realizable in nature. We then give the most general model equation which accommodates all these set-ups as special cases.

\subsection{Time-independent inertia}

First of all, as a reference, the special case of time-independent inertia is considered. This set-up is sketched in Figure~\ref{fig:set_ups}(a) (gray background).
The particle has a constant mass $m$ and a constant moment of inertia $J$. In this case, the Langevin equation of motion reads as
\begin{gather}
m \, \ddot{\vec{R}}(t) = \xi \, v_0 \, \uvec{n}(t)  - \xi \, \dot{\vec{R}}(t)  + \xi \sqrt{2 D} \, \vec{f}_{\rm st}(t), \label{eq:model_1_trans} \\
J \, \ddot{\phi}(t) = M - \xi_{r} \, \dot{\phi}(t) + \xi_{r} \sqrt{2 D_{r}} \, \tau_{\rm st}(t). \label{eq:model_1_rot}
\end{gather}
As far as the translational dynamics is concerned, there is a frictional damping force $- \xi \, \dot{\vec{R}}(t)$
and a self-propelling effective force along the particle orientation $\xi \, v_0 \, \uvec{n}(t)$ which gives rise to the particle self-propulsion velocity $v_0$ \cite{tenHagenWTKBL2015}. The latter does not stem from  mass ejection but is of another origin, such as diffusiophoresis or photophoresis.
This self-propulison force couples the orientational and translational degrees of freedom.
Furthermore there is a stochastic force (``noise'') $ \xi \sqrt{2D} \, \vec{f}_{\rm st}(t)$ where the effective translational diffusion coefficient $D$ quantifies the noise strength. We describe the stochastic term $\vec{f}_{\rm st}(t)$ as zero-mean Gaussian white noise with unit variance
\begin{equation}\label{eq:f_st}
\overline{ \vec{f}_{\rm st}(t) \otimes \vec{f}_{\rm st}(t')}= \delta(t-t') \id,
\end{equation}
where $\,\overline{\dotsb}\,$ indicates a noise average and $\id$ is the unit matrix.
Likewise, the rotational dynamics in Eq.~\eqref{eq:model_1_rot} involves a frictional torque $-\xi_{r} \, \dot{\phi}$  and an imposed torque $M$ plus the stochastic torque $\xi_{r} \sqrt{2D_{r}} \, \tau_{\rm st}(t)$ where the effective rotational diffusion coefficient $D_{r}$ now quantifies the rotational noise strength and the  Gaussian noise $\tau_{\rm st}(t)$ has  zero-mean and unit variance
\begin{equation}\label{eq:tau_st}
\overline{ \tau_{\rm st}(t) \tau_{\rm st}(t')} =\delta(t-t').
\end{equation}
One of the best experimental realization of active Langevin motion (see Eqs. \eqref{eq:model_1_trans} and \eqref{eq:model_1_rot}) can be found in self-propelled granular particles.
These particles are capable of transferring the energy of a vibrating surface or an internal motor to translational or rotational motion. 
Asymmetry in the particle design causes them to jump forward or to rotate when lifted from the ground.
From a recent experiment on these active granular particles \cite{ScholzJLL2018}, we list exemplary orders of magnitude for our model parameters $m=1~\text{g}$, $J=10~\text{g}\,\text{mm}^2$, $\xi=10~\text{g}/\text{s}$, $\xi_r=100~\text{g}~\text{mm}^2/\text{s}$, $D=100~\text{mm}^2/\text{s}$, $D_r=1/s$, $v_0=10-100~\text{mm}/\text{s}$ and $M=10^{-7}~\text{N}\,\text{m}$.
We shall revisit this standard situation again in Section~\ref{sec:constant_parameters}.
In the absence of any inertial effects, i.e. when $m=J=0$, the equations of motion are overdamped and lead to the standard picture of active Brownian motion \cite{HowseJRGVG2007,tenHagenTL2011,BechingerDLLRVV2016}.

\subsection{Directed  mass ejection}

A rocket is self-propelled by directed  mass ejection, so it establishes a fundamental set-up of time-dependent inertia.
In the typical geometry assumed here and shown in Figure~\ref{fig:set_ups}(b) (red background) the direction of the mass ejection is centrally outwards opposite to the particle orientation.
For simplicity, the mass ejection occurs with a constant velocity $u$ relative to the moving rocket ($u>0$) and the outlet coincides with the center of mass as indicated by a wedge in Fig.~\ref{fig:set_ups}(b). 
The general case where the ejection occurs not from the center but from a point distant to the center leads to additional terms which complicate the analysis and is left for future studies.

We assume, however, here for more generality that the rocket also has an internal motor, which leads to an additional self-propulsion of velocity $v_0$.
In typical descriptions of macroscopic rockets translational and rotational fluctuations are ignored.
While this is a reasonable assumption for macroscopic rockets, it breaks down for mini-rockets.
The characteristic equations of motion for a self-propelled particle with directed mass ejection are
\begin{align}
\frac{d}{dt} \Big( m(t) \, \dot{\vec{R}}(t) \Big) = & \xi \, v_0 \, \uvec{n}(t) - \xi \, \dot{\vec{R}}(t)  + \xi \sqrt{2 D} \, \vec{f}_{\rm st}(t) \nonumber \\
& - \dot{m}(t) (u \, \uvec{n}(t)-\dot{\vec{R}}(t)), \label{eq:model_2_trans} 
\end{align}
and the orientational equation of motion is given by \eqref{eq:model_1_rot}.

In discussing the basic physics of Eq.~\eqref{eq:model_2_trans}, we use Newton's second postulate which states that the total change in translational momentum is the total force which is in this case the sum of friction, translational stochastic and self-propulsion forces.
But even in the force-free case the ejected mass carries away the momentum
$\dot{m}(t) (u \, \uvec{n}(t) - \dot{\vec{R}}(t) )$
per unit time which needs to be included in the balance of \eqref{eq:model_2_trans} with a minus sign due to the conservation of total momentum, see also \cite{Sommerfeld1952,Thomson1966,PlastinoM1992}.
This constitutes in fact the thrust force which accelerates the rocket. 
It is important to note here that the special case of the traditional Tsiolkovsky rocket equation is obtained  as a special limit of no fluctuations, no frictions, no additional self-propulsion, no external torque and a vanishing initial angular velocity, i.e., for $D=D_r=\xi=\xi_r=v_0=M=\dot \phi(t=0)= 0$ \cite{Tsiolkovsky1903}.

Since we assume that the outlet/tank of the particle coincides with the center-of-mass, the moment of inertia is not affected by the mass ejection and remains constant. 
Hence the orientational motion is identical to the case of time-independent inertia.
Clearly, via the mass ejection, the two equations \eqref{eq:model_2_trans} and \eqref{eq:model_1_rot} are coupled.

Realizations of the rocket-like self-propelled objects can in principle be found for self-propelled Janus particles in a complex plasma which are laser-heated such that they evaporate mass in a certain direction \cite{KrasheninnikovPSS2010,NosenkoIM2010,NosenkoLKRZT2020} or even for inflated toy balloons \cite{Mueller,ManganD2015} or active granular particles equipped with compressed air tanks.
For the latter, we would expect the particle loss mass at a rate of approximately $\dot{m}=-1~\text{g}/\text{s}$ by exhausting air at a velocity $u= 100~\text{mm}/\text{s}$.	
Initial mass and moment of inertia are $m_0=10~\text{g}$ and $J=100~\text{g}\,\text{mm}^2$.
The remaining para\-meters are of the order of $\xi=10~\text{g}/\text{s}$, $\xi_r=100~\text{g}\,\text{mm}^2/\text{s}$, $D=100~\text{mm}^2/\text{s}$, $D_r=1/s$, $v_0=10-100~\text{mm}/\text{s}$ and $M=10^{-7}~\text{N}\,\text{m}$.
We finally remark that there is some overdamped counterpart of rocket-like motion in the osmoto-phoresis of semipermeable vesicles \cite{osmo2} where the ejection of molecules out of vesicle body leads to self-propulsion driven by the osmotic pressure difference \cite{osmo1} and for Janus-particles and nanorockets driven by reactive momentum-transfer \cite{EloulPFF2020,LiRW2016}.

\subsection{Isotropic mass evaporation}

A different situation occurs if the mass ejection is not directed but isotropic as sketched in Figure~\ref{fig:set_ups}(c) (green background).
Imagine a particle coated with an isotropic  layer that evaporates likewise in all directions, as realizable in dusty plasmas \cite{KrasheninnikovPSS2010,NosenkoIM2010,NosenkoLKRZT2020}.
In this case, the ejected mass does only carry away the translational momentum given by $- \dot{m}(t) \, \dot{\vec{R}}(t)$ such that the translational equations of motion for this case coincide with Eq.~\eqref{eq:model_2_trans} for $u=0$. However, the mass ejection is radial only in the body frame but for a rotating particle the angular momentum
${\dot J}(t) \, {\dot \phi}(t)$ is taken away in the laboratory frame even in the absence of any torque.
Therefore the orientational equation of motion reads as \eqref{eq:model_1_rot} with $J$ replaced by $J(t)$, as follows
\begin{equation}
J(t) \, \ddot{\phi}(t)     = M - \xi_{r} \, \dot{\phi}(t) + \xi_{r} \sqrt{2 D_{r}} \, \tau_{\rm st}(t)  .  \label{eq:model_3_rot}
\end{equation}
This set-up could be realized in experiments by placing a leaking water tank or evaporating material on an active granular particle.
The order of magnitude of the parameters might be $m_0=10~\text{g}$, $\dot{m}=-1~\text{g}/\text{s}$, $J_0=100~\text{g}\,\text{mm}^2$, $\xi=10~\text{g}/\text{s}$, $\xi_r=100~\text{g}\,\text{mm}^2/\text{s}$, $D=100~\text{mm}^2/\text{s}$, $D_r=1/s$, $v_0=10-100~\text{mm}/\text{s}$ and $M=10^{-7}~\text{N}\,\text{m}$.
Finally we remark that the inverse situation of mass adsorption can be treated in a similar way with a positive sign of $\dot{m}(t)$.

\subsection{Isotropic shape change}

The pirouette of figure skating  is an example of a  fourth situation where the total mass $m$ of the body is time-independent but the moment of inertia does change due to a shape change of the body.
In this special case,  sketched in Figure~\ref{fig:set_ups}(d) (blue background), the shape change does not carry away angular momentum but the total angular momentum is conserved.
Consequently while the translational equation of motion is identical with Eq.~\eqref{eq:model_1_trans}, the orientational equation of motion is given by
\begin{equation}
	\frac{d}{dt} \Big( J(t) \, \dot{\phi}(t) \Big)  =  M - \xi_{r} \, \dot{\phi}(t)  + \xi_{r} \sqrt{2 D_{r}} \, \tau_{\rm st}(t).  \label{eq:model_4_rot}
\end{equation}
Lastly this model could describe an active granular particle with a stretched elastic material attached to it.
In that way, the initial moment of inertia could be increased by an order of magnitude $J_0=100~\text{g}\,\text{mm}^2$, relaxing over a few seconds to its equilibrium shape with $\dot{J}=-1~\text{g}\,\text{mm}^2/\text{s}$.
The order of magnitude of the other parameters might be $m=1~\text{g}$, $\xi=10~\text{g}/\text{s}$, $\xi_r=100~\text{g}\,\text{mm}^2/\text{s}$, $D=100~\text{mm}^2/\text{s}$, $D_r=1/s$, $v_0=10-100~\text{mm}/\text{s}$ and $M=10^{-7}~\text{N}\,\text{m}$.

\subsection{General model}

The lesson to be learned from the previous examples is that the equations of motion depend on the imposed set-up of mass change.
In order to proceed in a general way, we now present a general framework of equations of motion which accommodates all previous special cases.
To define this model as general as possible, we also assume an effective time-dependent self-propulsion speed $v_0(t)$, a time-dependent internal torque  $M(t)$, a time-dependent translational $\xi(t)$ and rotational friction coefficient $\xi_r(t)$, as well as, time-dependent translational $D(t)$ and rotational diffusion coefficient $D_r(t)$ and a  time-dependent mass ejection velocity $u(t)$.

We now consider the following general  Langevin equations governing the  translational and the rotational motion for a self-propelled particle
\begin{align}
\frac{d}{dt} \Big( m(t) \, \dot{\vec{R}}(t) \Big) = & \xi(t) \big( v_0(t) \, \uvec{n}(t) - \dot{\vec{R}}(t)  + \sqrt{2 D(t)} \, \vec{f}_{\rm st}(t)  \big) \nonumber \\
& - \dot{m}(t)  \big( u(t) \, \uvec{n}(t) - \dot{\vec{R}}(t) \big), \label{eq:general_model_trans} \\
\frac{d}{dt} \Big( J(t) \, \dot{\phi}(t) \Big)  = &  M(t) - \xi_{r}(t) \, \dot{\phi}(t)  + \xi_{r}(t) \sqrt{2 D_{r}(t)} \, \tau_{\rm st}(t) \nonumber \\
& + \nu \, \dot{J}(t) \, \dot{\phi}(t). \label{eq:general_model_rot}
\end{align}
Clearly, all situations discussed so far and shown in Figure~\ref{fig:set_ups} are obtained from these equations as special cases: of course, \ref{fig:set_ups}(a) is the special limit where
the parameters $m$, $J$, $\xi$, $\xi_r$ $D$, $D_r$ $v_0$ and $M$ are constant. 
The rocket set-up in Figure~\ref{fig:set_ups}(b) coincides with the
general equations \eqref{eq:general_model_trans} and \eqref{eq:general_model_rot} when the parameters $J$, $\xi$, $\xi_r$ $D$, $D_r$ $v_0$, $M$, the relative velocity $u$ are constant.
The isotropic mass evaporation (Figure~\ref{fig:set_ups}(c)) is contained in \eqref{eq:general_model_trans} and \eqref{eq:general_model_rot} when 
the parameters $\xi$, $\xi_r$ $D$, $D_r$, $v_0$, $M$ are constant, the relative velocity vanishes $u=0$ and $\nu = 1$.
Finally the equations for an isotropic shape change (Figure~\ref{fig:set_ups}(d)) follow when in \eqref{eq:general_model_trans} and \eqref{eq:general_model_rot} the parameters $m$, $\xi$, $\xi_r$ $D$, $D_r$, $v_0$, $M$ are constant and $\nu = 0$.

At this stage we remark that more realistic situations can also be accommodated into the general equations \eqref{eq:general_model_trans} and \eqref{eq:general_model_rot}. 
These include, for example, a rocket where the outlet of the mass ejection does not coincide with the center-of-mass or where the ejection direction is not parallel to the particle orientation \cite{RankinR1949}.

From a mathematical point of view the equations of motion \eqref{eq:general_model_trans} and \eqref{eq:general_model_rot} are stochastic differential equations with Gaussian noise. The rotational equation \eqref{eq:general_model_rot}
is linear so that the distribution of the angle and angular velocity is Gaussian for any time. We give the corresponding
general solutions of \eqref{eq:general_model_trans} and \eqref{eq:general_model_rot}  in the Appendix \ref{sec:general_solution}.

\section{\label{sec:constant_parameters} Time-independent inertia}

We now turn to the special case of time-independent parameters defined by equations \eqref{eq:model_1_trans} and \eqref{eq:model_1_rot}. These equations of motion were studied before in references \cite{ScholzJLL2018,Loewen2020,Sandoval2020}.  
Here we summarize essential known results but also provide new analytical results for the full time resolved mean displacement, velocity correlation function and  mean-square displacement.
In doing so, we first consider the noise-free case and then include effects of noise.

\subsection{\label{sec:noNoise}Results for vanishing noise}

For given initial orientations $\phi_0 = \phi(0)$  and angular velocities $\dot{\phi}_{0}=\dot{\phi}(0)$ at time $t=0$, the
deterministic solution of the general  orientational equation of motion (2) in the absence of noise is
\begin{equation}\label{phi_NoNoise}
\phi(t) = \phi_{0} + \omega t + \frac{\dot{\phi}_{0} - \omega}{\gamma_{r}} \left(1 - e^{-\gamma_{r} t}\right).
\end{equation}
with the spinning frequency $\omega=M/\xi_r$ and rotational damping rate $\gamma_r=\xi_r/J$.
Plugging this solution into the noise-free translational equation (1), we obtain
for given initial positions $\vec{R}_0=\vec{R}(0)$   and velocities $ \dot{\vec{R}}_0 = \dot{\vec{R}}(0)$ at time $t=0$
the particle velocity
\begin{align}\label{noise_free_velocity}
\dot{\vec{R}}(t) = & \dot{\vec{R}}_{0}  e^{-\gamma t} + v_{0} \vec{\hat{P}}\Big[ \tilde{\gamma} \, \big( i \theta \big)^{\tilde{\gamma}+i\tilde{\omega}} e^{i( \phi_{0} + \theta )} \\
& \times \Gamma\big( -(\tilde{\gamma} + i\tilde{\omega}), i \theta e^{-\gamma_r t},  i \theta \big) \Big] e^{-\gamma t}, \nonumber
\end{align}
where  we introduced the translational damping rate  $\gamma=\xi/m$ and the notations
$\tilde{\gamma} = \gamma / \gamma_{r} $, $\tilde{\omega} = \omega/\gamma_{r}$, $\theta = (\dot{\phi}_{0} - \omega)/ \gamma_{r}$.
Moreover, $\Gamma(s,x_1,x_2)$ denotes the generalized Gamma function \cite{InGammaF},
\begin{equation}
\Gamma(s,x_1,x_2) = \integral{x_1}{x_2}{t} \, t^{s-1}e^{-t},
\end{equation}
and the operator $\vec{\hat{P}}$ formally transforms a complex number $z$ into its  two-dimensional vector $(\operatorname{Re} z, \operatorname{Im}z)$
in the complex plane.
This results in the particle trajectory
\begin{align} \label{noise_free_traj}
\vec{R}(t)  = & \vec{R}_0 + \frac{\dot{\vec{R}}_{0}}{\gamma}\big(1-e^{-\gamma t}\big) \\
& +\frac{v_{0}}{\gamma_{r}} \vec{\hat{P}}\Big[  \big( i \theta \big)^{i\tilde{\omega}} e^{i(\phi_{0}+\theta)} \Gamma\big( - i \tilde{\omega}, i \theta e^{-\gamma_r t},  i \theta \big) \Big]  \nonumber
\\
- \frac{v_{0}}{\gamma_{r}} \vec{\hat{P}}\Big[ & \big( i \theta \big)^{\tilde{\gamma}+i\tilde{\omega}}  e^{i(\phi_{0}+\theta)} \Gamma\big( -(\tilde{\gamma}+i\tilde{\omega}), i \theta e^{-\gamma_r t},  i \theta \big) \Big] e^{-\gamma t}  , \nonumber
\end{align}

In the limit of long times, the angular velocity reaches the spinning frequency,  $\lim_{t\to\infty} {\dot \phi}(t) = \omega$ so that
the particle is rotating with this frequency around a circle of radius
\begin{equation}
r =  \frac{v_0}{\omega} \sqrt{\frac{\gamma^2}{\gamma^2+\omega^2}},
\label{radius}
\end{equation}
centered at the position
\begin{equation}
\vec{R}_{c} = \vec{R}_{0}+ \frac{\dot{\vec{R}}_{0}}{\gamma}+\frac{v_{0}}{\gamma_{r}} \vec{\hat{P}}\Big[ \big( i \theta \big)^{i\tilde{\omega}}   e^{i(\phi_{0}+ \theta )} \Gamma\big( -i\tilde{\omega}, 0,  i \theta \big) \Big].
\end{equation}
Clearly, the spinning frequency $\omega$ does not depend on any inertia. However, the circle radius $r$ depends on the mass $m$ via the translational damping rate
$\gamma$ due to the centrifugal force but is independent on the moment of inertia $J$. The center of the circle depends on $\vec{R}_0$,  $\dot{\vec{R}}_0$, $\phi_0$,
$\dot{\phi}_{0}$, demonstrating that for vanishing  noise even the  long-time limit may depend on the initial conditions.
Finally, in the overdamped limit of vanishing inertia, the results reduce to that of Brownian circle swimmers \cite{vanTeeffelen2008,KurzthalerF2017}.

\subsection{\label{sec:Noisy}Effect of Brownian noise}

Subjected to Brownian noise, the particle will relax to a steady state after a long time forgetting about its initial conditions $\vec{R}_0$,  $\dot{\vec{R}}_0$, $\phi_0$, $\dot{\phi}_{0}$ at time $t=0$.
The static and dynamical correlation in the steady state can be  calculated as a time average over a very long time window which we shall denote with brackets $\mean{\dots}$. In the sequel we shall consider several of such dynamical correlations.
In the steady state, one can also calculate conditional averages.
For example one can built dynamical averages in the steady state after a lag time under the condition that the particle's position and orientation is prescribed at an initial time.
We shall compile analytical results for the different correlation functions first and show also examples for numerical evaluations of the analytical formula.

\subsubsection{Velocity correlation function}

First we introduce the translational velocity correlation function \cite{HansenMcDonald1990},
\begin{equation}
Z(t) = \mean{ \dot{\vec{R}}(t) \cdot \dot{\vec{R}}(0) },
\end{equation}
where $t$ now denotes a lag time and $\dot{\vec{R}}(0)$ is taken from the velocity  distribution in the steady state. 
We remark that the latter was computed recently for small inertia \cite{HerreraS2021} and for the formally equivalent model of an overdamped particle in a harmonic potential \cite{MalakarDKKD2020}.
The velocity  distribution is non-Gaussian (i.e.\ non Maxwellian) and its second moment $Z(0) = \mean{ \dot{\vec{R}}(0) \cdot \dot{\vec{R}}(0)}$ which is proportional to the the mean kinetic energy is known analytically \cite{ScholzJLL2018} as
\begin{equation} \label{vel_second_moment}
Z(0) = 2 D \gamma + v_{0}^{2}  \operatorname{Re}\Big[ \tilde{\gamma}e^{\tilde{D}_r} \tilde{D}_{r}^{-\Omega_{+}} \, \Gamma(\Omega_{+},0 ,\tilde{D}_r) \Big],
\end{equation}
where we introduced $\tilde D_r = D_r/\gamma_r$ and $\Omega_\pm  = \big(D_r \pm (\gamma + i \omega)\big) /\gamma_r$.
For an active inertial particle considered here, we have obtained the analytical result
\begin{equation} \label{vel_auto_corr}
Z(t) = 2D\gamma e^{-\gamma t}+ \frac{v_{0}}{2} \big( \mean{\dot{\vec{R}}(t)\cdot\uvec{n}(0)}  + \mean{\dot{\vec{R}}(0)\cdot\uvec{n}(t)} \big)
\end{equation}
with
\begin{alignat}{1} 
\mean{\dot{\vec{R}}(t)\cdot\uvec{n}(0)} = &  v_0 \operatorname{Re} \Big[ \tilde{\gamma} e^{\tilde{D}_r} \Big(  \tilde{D}_{r}^{-\Omega_{-}} \Gamma( \Omega_{-}, \tilde{D}_r e^{-\gamma_{r}t}, \tilde{D}_r) \nonumber \\
& \qquad \, \, \, \, + \tilde{D}_{r}^{-\Omega_{+}} \Gamma(\Omega_{+}, 0, \tilde{D}_r) \Big) e^{-\gamma t} \Big], \label{eq:V(t)n(0)}
\end{alignat}
and
\begin{equation} \label{eq:V(0)n(t)}
\mean{\dot{\vec{R}}(0)\cdot\uvec{n}(t)} =  v_0 \operatorname{Re} \Big[ \tilde{\gamma} e^{\tilde{D}_r} \tilde{D}_{r}^{-\Omega_{+}} \Gamma( \Omega_{+}, 0, \tilde{D}_r e^{-\gamma_{r}t}) e^{\gamma t} \Big].
\end{equation}
which implies that the long-time behavior of $Z(t)$ is exponential in time.

\subsubsection{Orientation correlation function}

Similarly the dynamical  orientational correlation function $C(t) = \mean{\uvec{n}(t)\cdot\uvec{n}(0)}$ in the steady state  can be expressed analytically as a double exponential as
\begin{align} \label{eq:orientcorr}
C(t) = \cos(\omega t) \, e^{-D_{r}(t-\gamma_r^{-1}(1-e^{-\gamma_r t}))},
\end{align}
which was found previously in another context by Ghosh and coworkers \cite{GhoshLMM2015} for $\omega=0$  and for general  $\omega$ in Ref.~\cite{ScholzJLL2018}.
Again it decays exponentially in time for long times.
A characteristic orientational persistence time $\tau_p$ can  be determined as
\begin{equation} \label{eq:persistence_time}
\tau_p = \int_{0}^{\infty} \! C(t) dt = \frac{1}{D_r} \operatorname{Re}\Big[\tilde{D}_r e^{\tilde{D}_r}\tilde{D}_{r}^{-\Omega} \, \Gamma(\Omega,0,\tilde{D}_r)\Big].
\end{equation}
with $\Omega = (D_r - i \omega) /\gamma_r$.
For vanishing inertia we recover the known result of the persistence time $\tau_p = D_r / (D_r^2 + \omega^2)$ \cite{vanTeeffelen2008, KurzthalerF2017} which simplifies further to the standard result $\tau_p = 1/D_r$ for linear swimmer \cite{HowseJRGVG2007}.

\subsubsection{\label{sec:mean_displacement} Mean displacement}

\begin{figure}[t]
	\centering
	\includegraphics[width = \columnwidth]{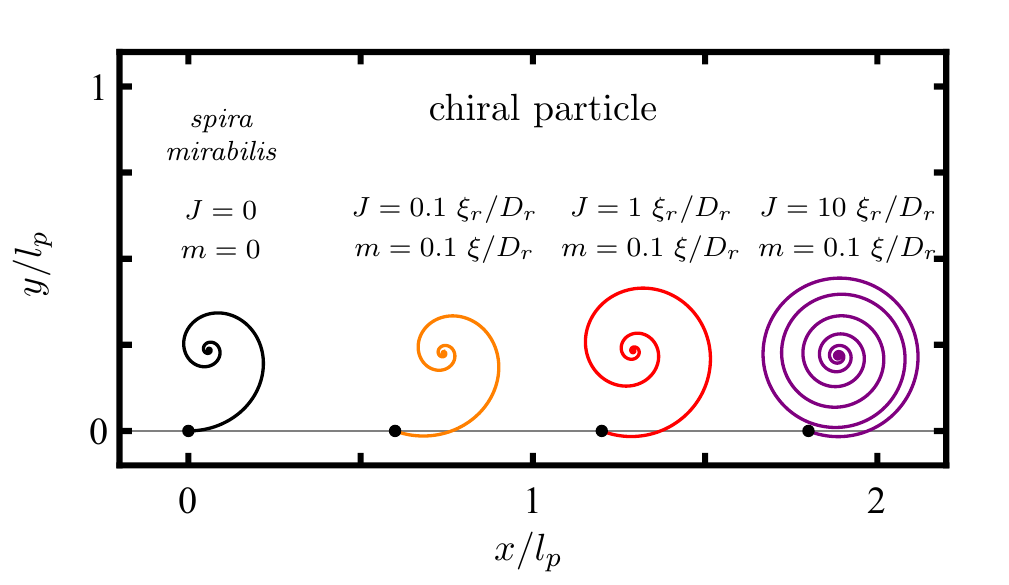}
	\caption{\label{fig:mean_displacement}
		Mean displacement $\mean{\Delta \vec{R}(t)}$ in the $xy$-plane for a chiral particle  with initial orientation along the $x$-axis for
		different moment of inertia, $J=0.1~\xi_r/D_r$ (orange), $J=1~\xi_r/D_r$ (red) and $J=10~\xi_r/D_r$ (purple). 
		Lengths are given in units of $l_p=v_0/D_r$. 
		The parameters are $\omega = 4~D_r$, $m=0.1~\xi/D_r$.
		The starting point at $t=0$ is denoted by a black dot.
		The \textit{spira mirabilis} of the  overdamped limit is plotted on the left (black) for comparison.
	}
\end{figure}

Next, we address the mean displacement $\mean{\Delta \vec{R}(t)} =  \mean{\vec{R}(t) - \vec{R}_0}$ of the particle in the steady state as a function of time $t$.
The average is now taken in the steady state but under the condition that for the initial time $t=0$, the position $\vec{R}(0)=\vec{R}_0$ and the orientation $\hat {\vec n}(0)$ (embodied in $\phi(0)=\phi_0$) are prescribed.
Since the particle velocities and the orientations are correlated in the steady state, the average over the translational velocity $\mean{\dot{\vec{R}}(0)}$ is not vanishing due to the prescribed orientation $\hat {\vec n}(0)$.
This average is given by
\begin{equation}
\mean{ \dot{\vec{R}}(0) } = v_{0} \vec{\hat{P}}\Big[ \tilde{\gamma} e^{\tilde{D}_r} \tilde{D}_{r}^{-\Omega_{+}} \Gamma(\Omega_{+}, 0, \tilde{D}_r) e^{i\phi_{0}} \Big].
\end{equation}

We obtain for the mean displacement
\begin{align} \label{eq:mean_displacement}
\mean{\Delta \vec{R}(t)}  = &  \frac{\mean{\dot{\vec{R}}(0)}}{\gamma} \Big( 1 - e^{-\gamma t} \Big) \\
&+ \frac{v_{0}}{D_{r}} \vec{\hat{P}}\Big[ \tilde{D}_r e^{\tilde{D}_r} \Big(  \tilde{D}_{r}^{-\Omega} \Gamma( \Omega, \tilde{D}_r e^{-\gamma_{r}t}, \tilde{D}_r) \nonumber \\
&+ \tilde{D}_{r}^{-\Omega_{-}} \Gamma( \Omega_{-}, \tilde{D}_r e^{-\gamma_{r}t}, \tilde{D}_r) e^{-\gamma t} \Big) e^{i\phi_{0}} \Big]. \nonumber
\end{align}
For short times $t$, the particle moves proceeds on average ballistically (i.e.\ linearly in time) with
\begin{equation}
\mean{ \Delta \vec{R}(t) } = \mean{\dot{\vec{R}}(0)} t + \mathcal{O}(t^2).
\label{MD_short_time_limit}
\end{equation}
Then the rotational noise decorrelates the current orientation from the  initial orientation and the mean displacement saturates to a finite persistence length $\vec{L}_p = \lim_{t\to\infty} \mean{ \Delta \vec{R}(t) }$ given as
\begin{equation} \label{MD_long_time_limit}
\vec{L}_p = \frac{\mean{\dot{\vec{R}}(0)}}{\gamma} + \frac{v_{0}}{D_{r}} \vec{\hat{P}}\Big[ \tilde{D}_r e^{\tilde{D}_r} \tilde{D}_{r}^{-\Omega} \, \Gamma(\Omega,0,\tilde{D}_r)  e^{i\phi_{0}} \Big].
\end{equation}
In case of a vanishing spinning frequency ($\omega =0$), the persistence length simplifies to $\vec{L}_p =  \mean{\dot{\vec{R}}(0)}/\gamma + v_{0} \tau_p \uvec{n}_0$
with $\tau_p$ given by (22).
In the overdamped limit we obtain the standard results of the persistence length for linear microswimmers $\vec{L}_p =   v_{0} \uvec{n}_0/D_r$ ($\omega =0$) \cite{HowseJRGVG2007}.
Moreover, for overdamped circle swimmer, the full time-resolved mean displacement given by \eqref{eq:mean_displacement} simplifies to a \textit{spira mirabilis} \cite{vanTeeffelen2008,KuemmeltHWBEVLB2013}. 
The presence of inertia will distort the ideal \textit{spira mirabilis} give rise to a more complex mean trajectory. 
This is shown in Figure~\ref{fig:mean_displacement} where three shapes of the mean trajectory for increasing moment of inertia $J$ are compared to the overdamped case.
Increasing $J$ reduces effectively the role of fluctuations such that there are more turns until the particle reaches half of the distance to  its final fixpoint.
Even though $\uvec{n}(0)$ is oriented towards the positive $x$-axis in all cases, the inertial mean trajectory first "oversteers" the initial orientation due to the velocity average, an effect which we shall elaborate and quantify further in subchapter 5.

\subsubsection{Mean-square displacement}

The full time-resolved mean-square-displacement (MSD) can be calculated as
\begin{equation}\label{MSD}
\mean{ \Delta \vec{R}^2(t) } = 4 D_L t  + \frac{2}{\gamma^2} \left( Z(t) - Z(0) \right) + 2\frac{v_{0}^{2}}{\gamma_r^{2}} F(t),
\end{equation}
with the long-time diffusion coefficient
\begin{equation} \label{eq:Long_time_diffusion_coefficient}
D_L = D+\frac{v_{0}^{2}}{2 D_r}\operatorname{Re}\Big[ \tilde{D}_r e^{\tilde{D}_r} \tilde{D}_{r}^{-\Omega} \, \Gamma(\Omega, 0, \tilde{D}_r) \Big],
\end{equation}
and the function
\begin{align}
F(t)  =  \Re \Bigg[&   \frac{e^{  \tilde{D}_r }}{\Omega^2} \Bigg(  \pFq{2}{2}{\Omega,\Omega}{\Omega+1,\Omega+1}{-  \tilde{D}_r }  \\
&-  \pFq{2}{2}{\Omega,\Omega}{\Omega+1,\Omega+1}{- \tilde{D}_r e^{- \gamma_{r} t }} e^{- \gamma_{r} \Omega t }  \Bigg) \Bigg], \nonumber
\end{align}
where ${}_pF_q$ represents the generalized hypergeometric function \cite{2F2}.

\begin{figure}[t]
	\centering
	\includegraphics[width = \columnwidth]{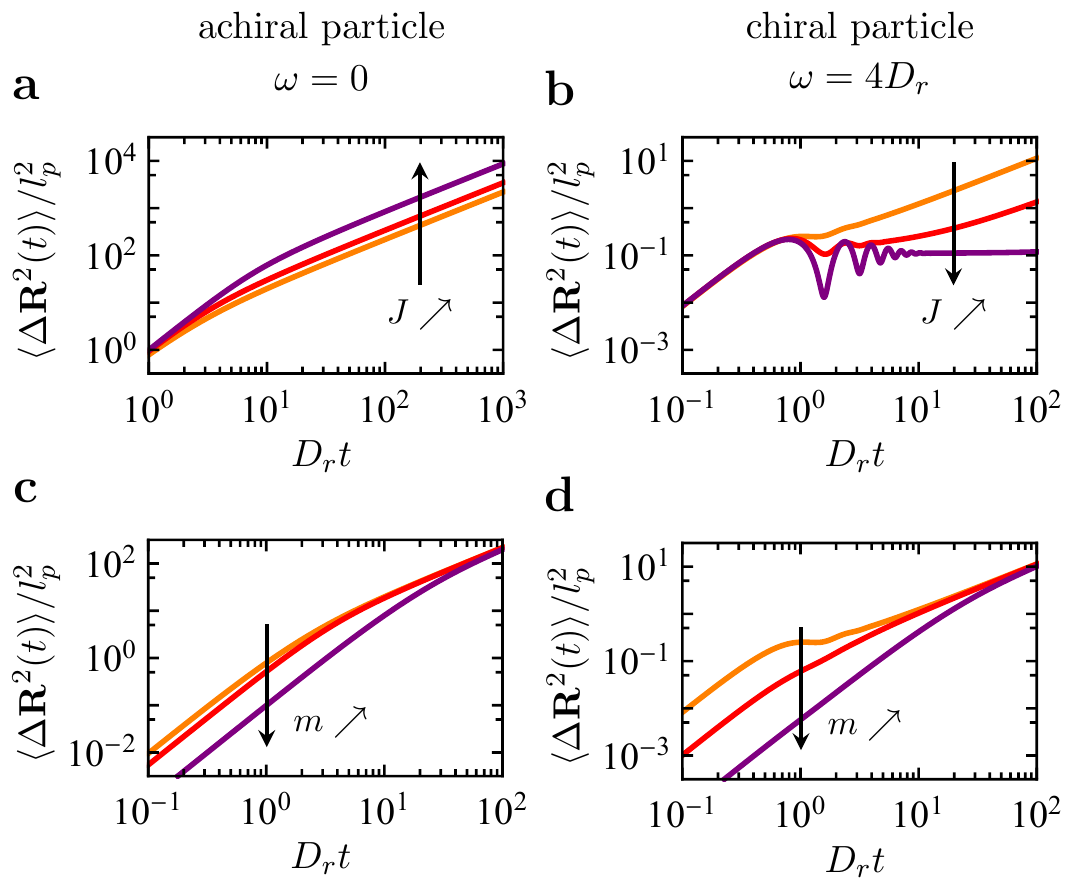}
	\caption{\label{fig:msd}
		Mean-square-displacement as a function of time on a double-logarithmic plot.
		(a) For an achiral particle with fixed mass $m$ and varied moment of inertia $J$.
		(b) For a chiral particle with fixed mass $m$ and varied moment of inertia $J$. 
		The fixed parameters are $m= 0.1~\xi/D_r$, $D=0$ and the moment of inertia is  $J=0.1~\xi_r/D_r$ (orange), $J=1~\xi_r/D_r$ (red) or $J=10~\xi_r/D_r$ (purple). 
		(c) For an achiral particle with varied $m$ and fixed $J$.
		(d) For a chiral particle with varied $m$ and fixed $J$.
		Here, the fixed parameters are $J= 0.1~\xi_r/D_r$, $D=0$ and the mass is $m=0.1~\xi/D_r$ (orange), $m=1~\xi/D_r$ (red) or $m=10~\xi/D_r$ (purple).
	}
\end{figure}

Figures~\ref{fig:msd}(a)-\ref{fig:msd}(d) compare the temporal behavior of the mean-square displacement of an achiral particle
to that of a chiral particle for different mases and moments of inertia $J$.
All curves exhibit the characteristic crossover from a short-time ballistic behavior
\begin{equation}
\mean{\Delta \vec{R}^2(t) } = Z(0) \, t^2 + \mathcal{O}(t^3),
\label{MSD_short_time_limit}
\end{equation}
to the long-time diffusive behavior governed by
\begin{equation}\label{eq:MSD_long_time_limit}
\mean{  \Delta\vec{R}^2(t) } \sim 4 D_L t.
\end{equation}
In the limit of small $J$, the short-time ballistic dynamics is
\begin{equation}
\lim_{J \to 0} \mean{ \Delta \vec{R}^2(t) }  = \Big( 2 D \gamma + v_{0}^{2} \, \frac{\gamma(\gamma+D_{r})}{(\gamma+D_{r})^{2}+\omega^{2}} \Big) t^2 + \mathcal{O}(t^3),
\label{MSD_short_time_limit_smallJ}
\end{equation}
while for large $J$, we have
\begin{equation}
\lim_{J \to \infty} \mean{ \Delta \vec{R}^2(t) } = \Big( 2 D \gamma +v_{0}^{2} \, \frac{\gamma^{2}}{\gamma^{2}+\omega^{2}} \Big) t^2 + \mathcal{O}(t^3).
\label{MSD_short_time_limit_bigJ}
\end{equation}
In general, the long-time diffusion coefficient $D_L$ (see Eq.~\eqref{eq:Long_time_diffusion_coefficient}) can be represented as
\begin{equation}
D_L = D + \frac{v_{0}^{2}}{2} \tau_\text{p},
\end{equation}
where the first term in Eq.~\eqref{eq:persistence_time} captures the diffusive behavior of a passive particle and the second is consistent with
the standard picture of a typical jump length of $v_0 \tau_p$ and a typical jump time of $\tau_p$, similar to the overdamped expression of microswimmers
when $\omega=0$ \cite{HowseJRGVG2007}.
It was emphasized in Ref.~\cite{ScholzJLL2018} that $D_L$ depends on the moment of inertia $J$ but not explicitly on the mass $m$.

In case of small moments of inertia, the long-time diffusion coefficient of the circle flyer asymptotically goes to \cite{ScholzJLL2018}
\begin{equation} \label{eq:Long_time_diffusion_coefficient_smallJ}
D_L = D + \frac{v_{0}^{2}}{2} \frac{D_r}{D_r^2+\omega^2} \left( 1 + \frac{D_r}{\xi_r} J \right) + \mathcal{O}(J^2),
\end{equation}
which grows dominantly proportional to moment of inertia.
The asymptotic behavior of the long-time diffusion coefficient for large moments of inertia is \cite{ScholzJLL2018},
\begin{equation} \label{eq:Long_time_diffusion_coefficient_bigJ}
D_L \sim
\begin{cases}
v_{0}^{2}\sqrt{\frac{\pi}{8D_{r}\xi_{r}}}\sqrt{J},  & (\omega = 0) \\
D, & (\omega \neq 0)
\end{cases}
\end{equation}
As the moment of inertia grows for $\omega\not=0$, the activity-induced part of the long-time diffusion coefficient goes asymptotically to zero (see Eq.~\eqref{eq:Long_time_diffusion_coefficient_bigJ}) since diffusion is hampered by systematic circling motion, i.e., the particle is get trapped in a circular path  due to its huge moment of inertia.

Figures~\ref{fig:msd}(a) and \ref{fig:msd}(c) show data for an achiral swimmer with different moments of inertia $J$ and different masses.
The short-time ballistic prefactor is pretty independent of $J$ but decreases with increasing $m$.
The latter trend follows from the fact that for large $m$ the particle cannot accelerate towards its self-propulsion velocity $v_0$.
Conversely, the long-time diffusivity is also increasing with $J$ according to \eqref{eq:Long_time_diffusion_coefficient_smallJ} and \eqref{eq:Long_time_diffusion_coefficient_bigJ} as the persistence in orientation increases with $J$ but it is independent of $m$. 
For a chiral particle, shown in Figures~\ref{fig:msd}(b) and \ref{fig:msd}(d), the MSD exhibits wiggles due to the circling.

\begin{figure}[t]
	\centering
	\includegraphics[width = 0.95\columnwidth]{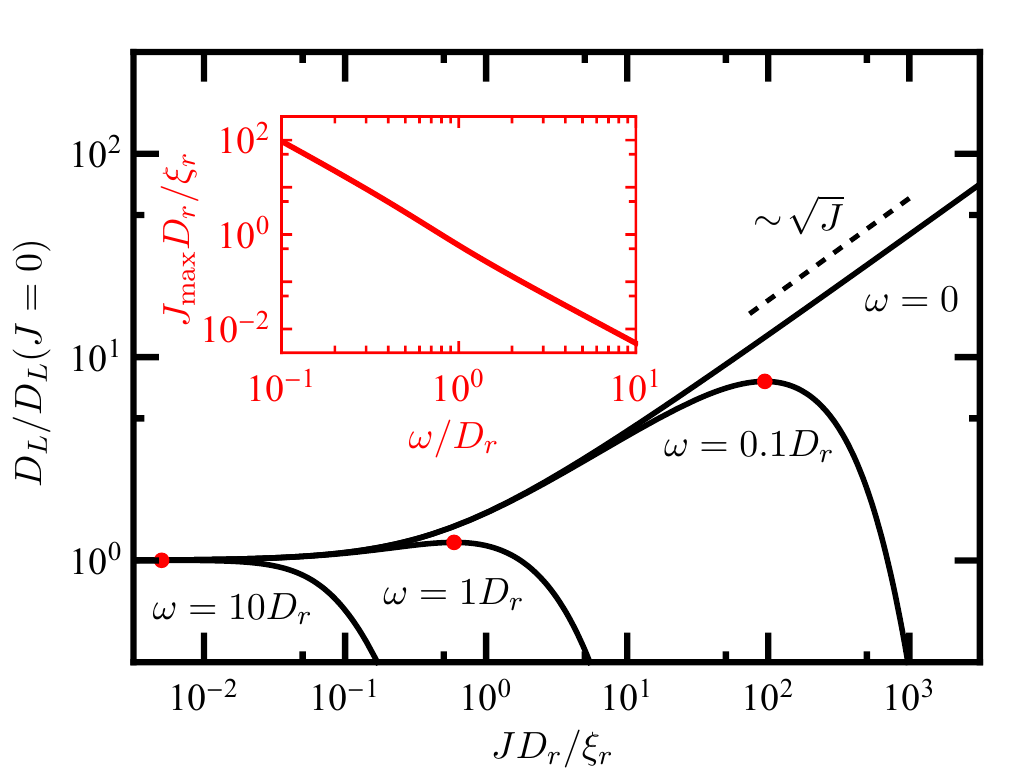}
	\caption{\label{Long_time_MSD_Fig}
		Long-time diffusion coefficient $D_L$ as a function of the moment of inertia $J$ for different circling frequencies $\omega= 10 D_r$, $\omega= 1 D_r$, $\omega= 0.1 D_r$ and $\omega= 0$. The translational diffusion coefficient was set to zero, $D=0$. In the inset, the  global maximum point $J_\text{max}$ for a given circling frequency $\omega$ is depicted. The corresponding maximal value  $D_L(J_\text{max})$ is shown as a red dot in the main figure.
	}
\end{figure}

An immediate consequence of \eqref{eq:Long_time_diffusion_coefficient_smallJ} and \eqref{eq:Long_time_diffusion_coefficient_bigJ} is that the long-time diffusivity behaves {\it non-monotonic\/} in $J$. 
Explicit data are presented in Figure \ref{Long_time_MSD_Fig} which illustrates the non-monotonic dependence of $D_L$ on the moment of inertia $J$ for different spinning frequencies $\omega$ for the special case $D=0$.  
There is an intermediate maximum in $D_L$ which is indicated in Figure~\ref{Long_time_MSD_Fig} by a red point.
This peak could  be exploited for an optimal exploration of an unknown territory by adapting the moment of inertia accordingly.
The associated optimal moment of inertia is plotted as a function of the spinning frequency $\omega$ in the inset of Figure~\ref{Long_time_MSD_Fig}.

\subsubsection{\label{sec:delay_function} Delay function}

Opposed to the overdamped case, the velocity of an inertial particle does not coincide with its self-propulsion direction and in Ref.~\cite{ScholzJLL2018} a dynamical correlation function, referred to as delay function $d(t)$, was introduced to quantify the delay between the velocity and orientation dynamics
\begin{equation} \label{vel_orient_corr}
d(t) = \mean{\dot{\vec{R}}(t)\cdot\uvec{n}(0) }-\langle\dot{\vec{R}}(0)\cdot \uvec{n}(t) \rangle .
\end{equation}
The "mixed" difference ensures that this function is trivially  zero in the overdamped limit but when non-zero its sign contains valuable information about the delay process between $\uvec{n}(t)$ and $\dot{\vec{R}}(t)$. 
If for example $d(t)$ is positive, this means that - on average - first the particle orientation changes and then the velocity will follow that change after a time $t$.
A positive $d(t)$ is the standard behavior exploited by the oversteering of racing cars which is also expected for achiral particles. 
The full analytical result for $d(t)$ directly follows from \eqref{eq:V(t)n(0)} and \eqref{eq:V(0)n(t)} and was given in Ref.~\cite{ScholzJLL2018}. 
Most notably, for an achiral particle, $d(t)$ has a positive peak after a typical delay time, while for a chiral particle, $d(t)$  oscillates due to the systematic change in orientation.
The latter oscillation was recently observed in macroscopic whirligig beetles swimming at the water surface \cite{Turner2019}.

Here we also provide analytical limits of small and large moments of inertia $J$. For small $J$ we get
\begin{equation} \label{delay_small_J}
	d(t)= 2 v_0 A(t) \left( 1+\frac{D_r}{\xi_r}J \right) + \mathcal{O}(J^2),
\end{equation}
with
\begin{alignat}{1} 
	A(t) = &  \frac{ \gamma \, D_r \, (\gamma^2-D_r^2-\omega^2) ( \cos(\omega t) \, e^{-D_r t}- e^{-\gamma t} ) }{\big((\gamma+D_r)^2+\omega^2\big)\big((\gamma-D_r)^2+\omega^2\big)} \nonumber \\
	& + \frac{ \gamma \, \omega \, (\gamma^2+D_r^2+\omega^2) \sin(\omega t) \, e^{-D_r t} }{\big((\gamma+D_r)^2+\omega^2\big)\big((\gamma-D_r)^2+\omega^2\big)}. 
\end{alignat}
Since $A(t)$ is positive for small times $t$, the delay effect is enhanced for increasing $J$.
Moreover the oscillatory behavior of chiral particle can be seen here directly.
In the opposite limit of large moment of inertia, the inertial delay approaches
\begin{equation}
\lim_{J\to \infty} d(t) = 2 v_0 \frac{\gamma\, \omega}{\gamma^2+\omega^2}\sin(\omega t),
\label{delay_limit_large_J}
\end{equation}
independent of the rotational diffusion coefficient $D_r$ which documents again the oscillatory behavior for chiral particles.

\section{\label{sec:time_dependent_inertia} Time-dependent inertia} 

Here we study the effect of time-dependent inertia on the Langevin motion of underdamped particle. 
We first introduce a reduced Langevin rocket model in which the mass of the particle gets burned to accelerate the particle giving rise to a time-dependent mass and propulsion speed.
Then we compare the four different set-ups introduced in chapter \ref{sec:theory}.
Last, we consider the limiting case of slowly varying parameters with respect to time. 

\subsection{\label{sec:rocket} Langevin rocket}

We define the ``Langevin rocket"
model by including orientational fluctuations in the  traditional Tsiolkovsky rocket equation \cite{Tsiolkovsky1903}. 
The effect of noise on rocket motion has been considered previously, see e.g.\ \cite{SrivastavaTK2012}, but a simple basis reference model for that is missing. 
We therefore simplify the general Eqs.~\eqref{eq:model_2_trans} and \eqref{eq:model_1_rot} for directed mass ejection and
assume vanishing moment of inertia, torque and translational diffusion ($J=0$, $M=0$ and $D=0$)
The Langevin rocket dynamics for a prescribed $m(t)$ is then given by
\begin{gather}
m(t) \, \ddot{\vec{R}}(t) + \xi \, \dot{\vec{R}}(t) = - u \, \dot{m}(t) \, \vec{n}(t),
\label{eq:R(t)_langevin_rocket} \\
\dot{\phi} (t)= \sqrt{2D_{r}} \, \tau_{\rm st}(t).
\label{eq:phi(t)_langevin_rocket}
\end{gather}
This set of equations approaches the ideal Tsiolkovsky rocket equation, $m(t) \, \ddot{\vec{R}}(t)  = - u \, \dot{m}(t) \, \vec{n}_0$, in the limit of vanishing damping $(\xi=0)$ and noise ($D_r=0$) \cite{Tsiolkovsky1903}.

For the sake of simplicity, we assume that the rocket is ejecting mass at a constant rate $( m_\infty - m_0 )/\Delta t$ where
$m_0$ denotes the initial mass, $m_\infty$ the final rest mass of the rocket and $\Delta t$ is the total burn time. The ejection process happens in the window
$0<t<\Delta t$ such that
the time dependent particle mass is
\begin{equation} \label{eq:mass_linear_decay}
m(t) = m_0 + ( m_\infty - m_0 ) \frac{\min(t,\Delta t)}{\Delta t}.
\end{equation}
In the following, we discuss the average reach of the rocket (i.e.\ its mean displacement)
as a function time. In particular, we investigate the final reach for long times as a
function of the burn time $\Delta t$ and the propellant mass fraction
\begin{equation} \label{eq:propellant_mass_fraction}
\zeta = \frac{m_0 - m_\infty}{m_0}.
\end{equation}

\subsubsection{Results for vanishing noise}

In the absence of rotational noise, the displacement of the rocket for a vanishing initial velocity at $t=0$ is
{\allowdisplaybreaks
	\begin{align}
	&\Delta \vec{R}(t) =  u  \frac{\min(t,\Delta t)}{S_1 + 1} \uvec{n}_0 - \frac{u}{\gamma_0} \frac{m(t)}{m_0}  \frac{1 - \big( \smallfrac{m(t)}{m_0} \big)^{S_1}}{S_1 + 1} \uvec{n}_0 \\
	&+ \frac{u}{\gamma_0} \frac{m(t)}{m_0} \frac{ 1 - \big( \smallfrac{m(t)}{m_0} \big)^{S_1} }{S_1} \Bigg( 1 - e^{- \gamma_\infty (\max(t,\Delta t) - \Delta t) } \Bigg) \uvec{n}_0 \nonumber,
	\end{align}}
with the initial damping rate  $\gamma_0= \xi/m_0$, the final damping rate $\gamma_\infty= \xi/m_\infty$ and the reduced burn time $S_1 = \gamma_0 \Delta t / \zeta$.

\begin{figure}[t]
	\centering
	\includegraphics[width = 0.9\columnwidth]{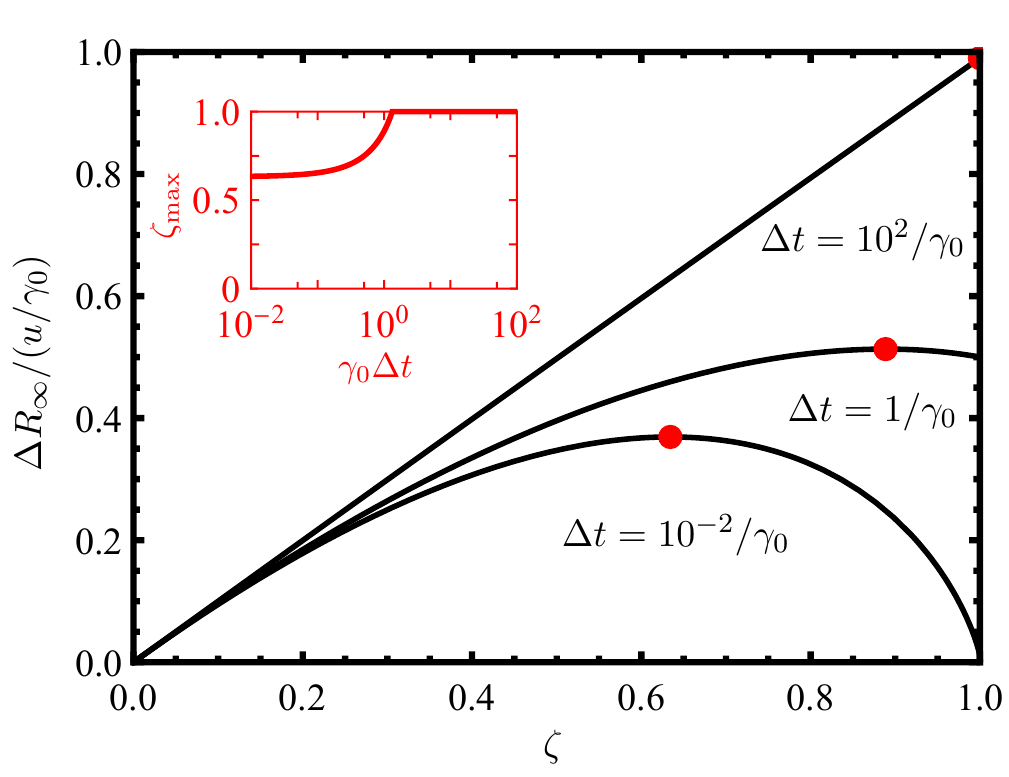}
	\caption{\label{fig:long_time_reach}
		Maximal reach $\Delta R_\infty$ as a function of the propellant mass fraction $\zeta$ for several burn times $\Delta t=10^{-2} /\gamma_0$,  $\Delta t=1 /\gamma_0$ and  $\Delta t=10^{2} /\gamma_0$.
		The inset depicts the  optimal propellant mass fraction $\zeta_\text{max}$ for a given burn time $\Delta t$. The corresponding maximal value  $R_{\infty}(\zeta_\text{max})$  is shown as a red dot in the main figure.
	}
\end{figure}

For short times, the rocket exhibits an acceleration by ejecting mass such that the displacement scales with $t^2$,
\begin{equation} \label{eq:rocket_short time}
\Delta \vec{R}(t) = \frac{u \zeta}{2 \Delta t} \uvec{n}_0 \, t^2 + \O{t^3}.
\end{equation}
After the burn time $\Delta t$, the rocket reaches its maximal velocity which is subsequently exponentially damped with the final damping rate $\gamma_\infty$ until the rocket comes to a standstill.
The total long-time displacement $\Delta R_\infty = \lim_{t\to\infty}  \abs{ \Delta \vec{R}(t) }$ is given by
\begin{equation}
\Delta R_\infty =  \frac{ u \Delta t}{S_1 + 1} +
\frac{u}{\gamma_0} \frac{  (1-\zeta)  \big(1 - (1-\zeta)^{S_1} \big) }{S_1 (S_1 + 1)}.
\end{equation}
In Figure~\ref{fig:long_time_reach}, we show the long-time displacement $\Delta R_\infty$ as a function of the propellant mass fraction $\zeta$ for different burn times $\Delta t$.
For long burn times $\Delta t \gg 1/\gamma_0$, the ultimate displacement increases linearly with the propellant mass fraction $\Delta R_\infty \sim u \, \zeta / \gamma_0$.
The rocket reaches the longest distance when $\zeta_\text{max} \sim 1$, a situation which can be called {\it complete extended mass ejection}.
Interestingly, however, there is a qualitative different behavior for burn times which are comparable or smaller than the characteristic damping time $1/\gamma_0$ where the displacements behave non-monotonic in the mass fraction $\zeta$. 
This can be intuitively understood as follows: for small mass fractions, more ejection means more propulsion and acceleration such that $\Delta R_\infty$ increases with $\zeta$.
Conversely for $\zeta$ close to 1, the rocket becomes very light after the burn time and therefore very quickly stops within an extremely short damping time $1/\gamma_\infty$ which reduces its reach relative to a situation of smaller $\zeta$.
Consequently the optimal value $\zeta_\text{max}$ for the mass ratio  for which the reach is maximal is smaller than $1$.
These corresponding optimal mass ratios are marked by red points in Figure~\ref{fig:long_time_reach} and plotted  as a function of the reduced burn time in the inset. 
For decreasing burn times the optimal mass ratio $\zeta_\text{max}$ exhibits a bifurcation-like behavior from complete mass ejection to a finite fraction with the special limit of $\zeta_\text{max} \sim 1 - e^{-1} \approx 0.63$ as $\Delta t$ approaches zero.

The special limit of $\Delta t \ll 1/\gamma_0$ deserves some more attention. In this case of {\it fractional instantaneous mass ejection},
the particle ejects only a fraction of its propellant to gain momentum very quickly. But it keeps a rest mass in order
to still proceed during the subsequent damping time.  In this limit we obtain
\begin{equation}\label{eq:rocket_trajectory_0}
\Delta \vec{R}(t) = - \frac{u}{\gamma_0} (1 - \zeta) \ln\big( 1 - \zeta \big) \left( 1 - e^{-\gamma_\infty t} \right) \uvec{n}_0,
\end{equation}
which scales for $\Delta t \ll t \ll 1/\gamma_0$  linearly in time
\begin{equation} \label{eq:rocket_short time_0}
\Delta \vec{R}(t) =- u  \, \ln \big( 1 - \zeta \big) \, \uvec{n}_0 \, t + \O{t^2},
\end{equation}
For long times, $t \gg 1/\gamma_0$, we obtain
\begin{equation}
\Delta R_\infty = - \frac{u}{\gamma_0}(1 - \zeta) \ln(1-\zeta).
\end{equation}

We finally remark that one can consider a full {\it optimization problem} with respect to both burn time $\Delta t$ and the mass fraction $\zeta$ by
posing the question: what is the maximal reach of the rocket
if the burn time $\Delta t$ and the mass fraction $\zeta$ can be varied freely and independently? The answer in the fluctuation-free case
is simple: the best strategy is to burn all mass $\zeta_\text{max} \to 1$ and do this over a very long time $\Delta t \to \infty$.
Then one achieves the maximum
\begin{equation}\label{eq:max_reach}
\max (\Delta R_\infty)=\frac{u}{\gamma_0},
\end{equation}
in the upper right corner of
Figure \ref{fig:long_time_reach}. In other terms, the strategy of  complete extended mass ejection  always outperforms that of
an fractional  instantaneous mass ejection.
This simple answer will change if orientational noise is included, a case which we shall address next.

\subsubsection{Noise-averaged mean reach and noise-induced transition between two mass ejection strategies}

\begin{figure*}[t]
	\centering
	\includegraphics[width = 2.\columnwidth]{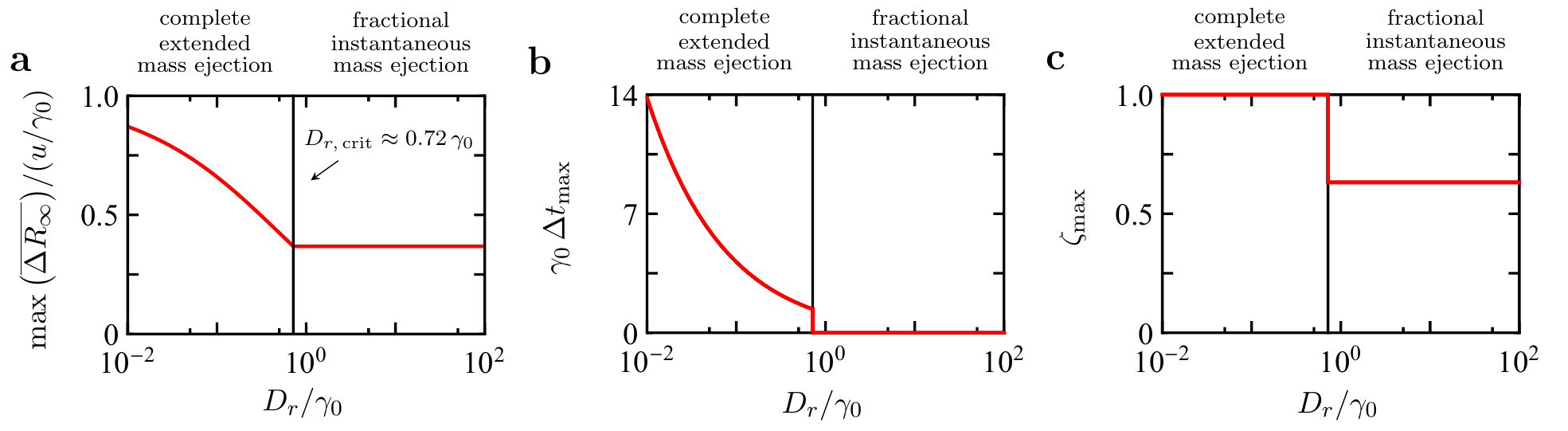}
	\caption{\label{fig:long_time_reach_max}
		(a) The optimal mean reach $\max \big( \overline{\Delta R_{\infty}} \big)$ maximized with respect to the propellant mass fraction $\zeta$ and the burn time $\Delta t$ as a function of the rotational noise strength $D_r/\gamma_0$.
		(b) Optimal burn time $\Delta t_\text{max}$ as a function of the rotational noise strength $D_r/\gamma_0$.
		(c) Optimal propellant mass fraction $\zeta_\text{max}$ as a function of the rotational noise strength $D_r/\gamma_0$.
		The transition from the complete extended mass ejection strategy to that of the fractional instantaneous mass ejection
		is marked by vertical black lines at $D_{r, \, \text{crit}} \approx 0.72 \, \gamma_0$ in all three figures.
	}
\end{figure*}

In case of finite rotational noise $(D_r>0$), we obtain for the noise-averaged displacement of the Langevin rocket the analytical result
\begin{align}
&\overline{\Delta \vec{R}(t)} = \frac{u}{D_r}  \frac{1}{S_1 + 1} \left( 1 - e^{- D_r \min(t,\Delta t)  } \right) \uvec{n}_0 \\
& + \frac{u}{D_r} \Re\bigg[  e^{S_2}  \big( - S_2 \big)^{S_1 + 1} \Gamma\Big( - S_1,- S_2 \big(\smallfrac{m(t)}{m_0}\big), - S_2  \Big) \bigg] \uvec{n}_0 \nonumber \\
& \times \bigg( \frac{ \big( \smallfrac{m(t)}{m_0} \big) ^{S_1+1}}{S_1 + 1} - \frac{ \big( \smallfrac{m(t)}{m_0} \big)^{S_1+1}}{S_1} \left( 1 - e^{- \gamma_\infty (\max(t,\Delta t) - \Delta t) } \right) \bigg), \nonumber 
\end{align}
with $S_2 = D_r \Delta t / \zeta $ proportional to the rotational noise.
Orientational fluctuations do not contribute to the short-time behavior as witnessed by the fact that
in this limit  the mean displacement coincides with the noise-free acceleration behavior of  Eq.~\eqref{eq:rocket_short time}.
For long times, on the other hand, the mean reach of  the Langevin rocket is
\begin{align} \label{eq:rocket_mean_long_time_reach}
&\overline{\Delta R_\infty} = \frac{u}{D_r} \frac{1}{S_1 + 1} \left( 1 - e^{- D_r \Delta t } \right) + \frac{u}{D_r} \frac{  (1 - \zeta)^{S_1 + 1} }{S_1 (S_1 + 1) }  \nonumber \\
& \times \Re\Big[  e^{S_2}  \big( - S_2 \big)^{S_1 + 1} \Gamma\big( - S_1,- S_2 (1 - \zeta), - S_2  \big) \Big]. 
\end{align}
Returning to the previous  optimization problem, we now maximize  the mean reach as a function of  burn time $\Delta t$ and mass fraction $\zeta$ for fixed prescribed noise strength $D_r/\gamma_0$.
In Figure \ref{fig:long_time_reach_max}(a), the resulting maximal reach $\max( \overline{\Delta R_\infty} )$ is shown for varied noise strength $D_r/\gamma_0$ in units of its universal noise-free limit $u/\gamma_0$ of complete extended mass ejection. 
The associated  optimal burn time $\Delta t_\text{max}$ and optimal mass fraction $\zeta_\text{max}$ are also presented (see Figure~\ref{fig:long_time_reach_max}(b) and \ref{fig:long_time_reach_max}(c)).
If rotational noise is increased, the complete extended mass ejection is still the best strategy, but it is optimal to burn the full mass over a finite burn time. 
This strategy defies best the ultimate orientational decorrelation which reduces the mean reach.
In the opposite limit of very large orientational noise, the best strategy is to get momentum quickly by ejecting  apart of the mass and use it to proceed further within the characteristic damping time. 
If one would eject the mass completely, the system is overdamped  after the burn time and will stop immediately lacking the additional benefit of the inertia.
Hence the fractional instantaneous mass ejection is the optimal strategy. 
Interestingly, there is a sharp noise-induced discontinuous transition between the two strategies for an intermediate finite value
\begin{gather}
D_{r, \text{crit}} \approx 0.72 \, \gamma_0,
\end{gather}
of the orientational noise. 
The latter is signaled by a sharp jump in the optimal burn time from  $1.39 \, \gamma_0$ to $0$ (see Figure~\ref{fig:long_time_reach_max}(b)). 
The optimal propellant mass fraction jumps from 1 to the universal value of $1-e^{-1} \approx 0.63$ (see Figure~\ref{fig:long_time_reach_max}(c)) and can thus be viewed as the "order parameter" of the transition.

\subsection{\label{sec:setups} Comparison between the different set-ups} 

We now compare the different set-ups for time-dependent inertia as discussed in Section~\ref{sec:theory} in more detail (see again Figure~\ref{fig:set_ups}). In the case of directed mass ejection or isotropic mass evaporation (Figure~\ref{fig:set_ups}(b) and \ref{fig:set_ups}(c)), we assume a mass loss exponentially in time $t$ as
\begin{equation} \label{eq:m(t)_decay_exp}
m(t) = m_\infty + ( m_0 - m_\infty ) e^{-\gamma_{m} t} ,
\end{equation}
where $m_0$ is the initial mass, $m_\infty$ the rest mass which remains after the fuel is burned and $\gamma_{m}$ the mass decay rate.
As outlined in Appendix~\ref{sec:C}, an exponential mass loss occurs in particular for a rocket which ejects gas molecules at constant speed from a tank under isothermal and isochoric conditions. 
In this case the exponential mass reduction follows from  the reduction of the gas density in the tank.
Accordingly we also assume an exponential decrease in the moment of inertia
\begin{equation} \label{eq:J(t)_decay_exp}
J(t) = J_\infty + ( J_0 - J_\infty ) e^{-\gamma_{J} t} ,
\end{equation}
where $J_0$ is the initial and $J_\infty$ the final moment of inertia and $\gamma_{J}$ the decay rate of the moment of inertia.
For the isotropic shape change (Figure~\ref{fig:set_ups}(d))
the mass is assumed to be constant and only an exponential loss in the moment of inertia is prescribed.

\begin{table*} [t]
	\begin{ruledtabular}
		\begin{tabular}{l@{\hspace{0.5cm}}|c|c|c|c}
			& time-independent inertia & directed mass ejection & isotropic mass ejection & isotropic shape change \\
			\hline
			$m(t)$ & $m_0$ &  $m_\infty + (m_0 - m_\infty) e^{-\gamma_m t} $ &  $ m_\infty + (m_0 - m_\infty) e^{-\gamma_m t} $  &  $m_0$ \\
			$\gamma_m$ & - & 0.1~$D_r$ & 0.1~$D_r$ & - \\
			$m_\infty / m_0$ & - & 0.1 & 0.1 & - \\
			$u$ & - & 1~$v_0$ & - & - \\
			$J(t)$ & $J_0$ &  $J_0$ &  $J_\infty + (J_0 - J_\infty) e^{-\gamma_J t} $  &  $J_\infty + (J_0 - J_\infty) e^{-\gamma_J t} $ \\
			$\gamma_J$ & - & - & 0.1~$D_r$ & 0.1~$D_r$ \\
			$J_\infty / J_0$ & - & - & 0.1 & 0.1 
		\end{tabular}
	\end{ruledtabular}
	\caption{\label{table:parameter} Simulation parameter for the different set-ups. 
	}
\end{table*}

The protocol is as follows. At time $t=0$, we start from a steady state achieved for constant parameters and then initiate the mass loss and moment of inertia change (or in general arbitrary time-dependencies).
For the different dynamical correlation functions we correlate the system configuration after a time $t$ with the steady state condition at time $t=0$ (over which we perform the average).
For the different dynamical correlation functions we correlate the steady state condition at time $t=0$ over which we perform the average
with the system configuration after a time $t$.
Under these conditions,  we obtain general analytical results for arbitrary time-dependencies. Since the system is relaxing or "aging", the two-point correlation functions now depend explicitly on two times $t_1$, $t_2$, not just on the time difference as in the steady state. 

For $t_1<t_2$, the orientational correlation function $C(t_1,t_2) = \mean{ \uvec{n}(t_1) \cdot \uvec{n}(t_2) }$ is given by
\begin{equation} \label{eq:generall_orientcorr}
C(t_1,t_2) = \cos\big( \mu(t_1, t_2)  \big) \, e^{- \smallfrac{1}{2} \sigma(t_1, t_2)},
\end{equation}
with the mean angle difference
\begin{equation}
\mu(t_1, t_2) = \integral{t_1}{t_2}{t''} \integral{-\infty}{t''}{t'} \, \frac{\xi_r(t')}{J(t')} \, \omega(t')  e^{-\Gamma_r(t',t'')},
\end{equation}
the corresponding variance 
\begin{align}
&\sigma(t_1, t_2) = 4 \integral{t_1}{t_2}{t'''} \integral{t_1}{t'''}{t''}\\
& \quad \times   \bigg( \integral{-\infty}{t''}{t'}  \bigg(\frac{\xi_r(t')}{J(t')}\bigg)^2  D_{r}(t') e^{-2\Gamma_r(t',t'')} \bigg) e^{-\Gamma_r(t'',t''')}. \nonumber
\end{align} 
and the rotational damping function 
\begin{equation}
\Gamma_r(t_1,t_2) = \int_{t_1}^{t_2} dt' \, \frac{\xi_r(t')}{J(t')} + (1 - \nu) \ln \left( \frac{J(t_2)}{J(t_1)} \right).
\end{equation}
Here, $\nu = 0$ in the case of isotropic shape change and $\nu = 1$ in the case of isotropic mass evaporation.

Similarly, the velocity correlation function $Z(t_1,t_2) = \mean{ \dot{\vec{R}}(t_1) \cdot \dot{\vec{R}}(t_2) }$ for $t_1<t_2$ is
\begin{align}
& Z(t_1,t_2) = 4 \int_{-\infty}^{t_1}\hspace{-0.1cm}\dif{t'}  \bigg(\frac{\xi(t')}{m(t')}\bigg)^2  D(t')  e^{-\Gamma(t',t_1)} e^{-\Gamma(t',t_2)}  \\
& + \int_{-\infty}^{t_1}\hspace{-0.25cm}\dif{t'}  \int_{-\infty}^{t_2}\hspace{-0.25cm}\dif{t''} a(t') a(t'')  \mean{ \uvec{n}(t') \cdot \uvec{n}(t'') }   e^{-\Gamma(t',t_1)}  e^{-\Gamma(t'',t_2)}, \nonumber
\end{align}
with the acceleration 
\begin{equation}
a(t) = \frac{\xi(t)}{m(t)} v_{0}(t) - \frac{\dot{m}(t)}{m(t)} u(t),
\end{equation}
and the translational damping function
\begin{equation}
\Gamma(t_1,t_2) = \int_{t_1}^{t_2} dt' \, \frac{\xi(t')}{m(t')}. 
\end{equation}

For the delay function $d(t_1,t_2) = \mean{\dot{\vec{R}}(t_2)\cdot\uvec{n}(t_1) }-\langle\dot{\vec{R}}(t_1)\cdot \uvec{n}(t_2) \rangle$, we obtain 
\begin{align}
d(t_1,t_2) = & \integral{-\infty}{t_2}{t'} a(t') \, \mean{ \uvec{n}(t') \cdot \uvec{n}(t_1) } \,  e^{-\Gamma(t',t_2)} \\
&- \integral{-\infty}{t_1}{t'} a(t') \, \mean{ \uvec{n}(t') \cdot \uvec{n}(t_2) }  \,  e^{-\Gamma(t',t_1)}. \nonumber
\end{align}
The general expression for the mean displacement $\mean{ \Delta \vec{R}(t_1,t_2)} = \mean{ \vec{R}(t_2) - \vec{R}(t_1) } $ is
\begin{equation}
\mean{ \Delta \vec{R}(t_1,t_2)}= \integral{t_1}{t_2}{t'} \integral{-\infty}{t'}{t''} a(t'') \, \mean{\uvec{n}(t'') \vert \uvec{n}(t_1) } e^{-\Gamma(t'',t')},
\end{equation}
where the conditional average 
\begin{align}
&\mean{ \uvec{n}(t_2) \vert \uvec{n}(t_1) } \\
&= \begin{cases}
\uvec{P} \Big[ e^{- \smallfrac{1}{2} \sigma(t_2, t_1) + i (\phi_{1} + \mu(t_2, t_1) ) } \Big], & \text{for} \, t_2>t_1, \\
\uvec{P} \Big[ e^{- \smallfrac{1}{2} \sigma(t_1, t_2) + i (\phi_{1} + \mu(t_1, t_2) ) } \Big], & \text{for} \, t_2<t_1,
\end{cases} \nonumber
\end{align}
denotes the mean orientation under the condition that the particle has the angle $\phi(t_1) = \phi_{1}$ at time $t_1$.

Last, the mean-square displacement $\mean{ \Delta \vec{R}^2(t_1,t_2)} = \mean{ (\vec{R}(t_2) - \vec{R}(t_1))^2} $ is 
\begin{equation}
\mean{\Delta \vec{R}^2(t_1,t_2)}  = \integral{t_1}{t_2}{t'}  \integral{t_1}{t_2}{t''} Z(t',t'').
\end{equation}

For time-independent parameters, we recover the results discussed in section \ref{sec:constant_parameters}. 
In particular, we have $C(t_1,t_2) = C(\abs{t_1 - t_2})$  (see Eq.~\eqref{eq:orientcorr}),  $Z(t_1,t_2) = Z(\abs{t_1 - t_2})$ (see Eq.~\eqref{vel_auto_corr}), $d(t_1,t_2) = d(\abs{t_1 - t_2})$ (see Eq.~\eqref{vel_orient_corr}), 
$\mean{\Delta \vec{R}(t_1,t_2)} = \mean{ \Delta \vec{R}(\abs{t_1 - t_2})}$ (see Eq.~\eqref{eq:mean_displacement}) and $\mean{\Delta \vec{R}^2(t_1,t_2)} = \mean{ \Delta \vec{R}^2(\abs{t_1 - t_2})}$ (see Eq.~\eqref{MSD}).

\begin{figure}[t]
	\centering
	\includegraphics[width = \columnwidth]{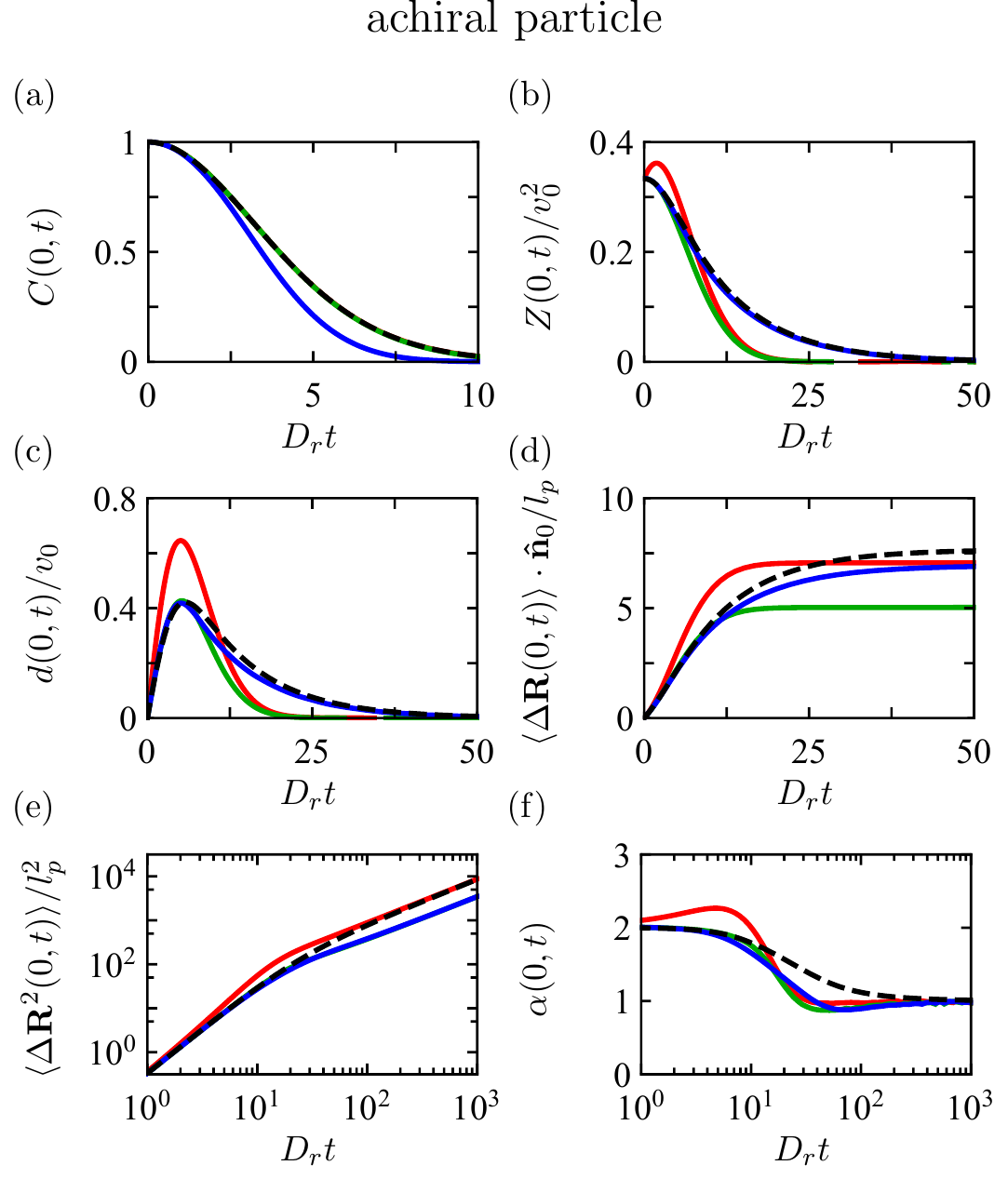}
	\caption{\label{fig:inertia_achiral}  
		Comparison of the different special set-ups for an achiral active particle ($\omega= 0$) with time-dependent inertia:
		(a) orientation autocorrelation function $C(0,t)$, 
		(b) velocity autocorrelation function $Z(0,t)$,
		(c) delay function $d(0,t)$,
		(d) mean displacement along the initial orientation $\mean{\Delta \vec{R}(0,t)} \cdot \uvec{n}_0$,
		(e) mean-square displacement $\mean{\Delta\vec{R}^2(0,t)}$ and (f) the corresponding scaling exponent $\alpha(0,t)$ for time-independent inertia (dashed black), directed mass ejection (red), isotropic mass evaporation (green) and an isotropic change in the particle shape (blue).
		Velocities are given in units of $v_0$, times in $1/D_r$ and lengths in $l_p=v_0/D_r$. 
		The time-dependencies of the mass $m(t)$ and the moment of inertia $J(t)$ for the different set-ups are summarized in Table~\ref{table:parameter}.
		The remaining parameters are $D= 0$, $\gamma_0 = \xi / m_0 = 0.1 D_r$ and $\gamma_{r,0} = \xi_{r} / J_0 = 0.1 D_r$.
	}
\end{figure}

\begin{figure}[t]
	\centering
	\includegraphics[width = \columnwidth]{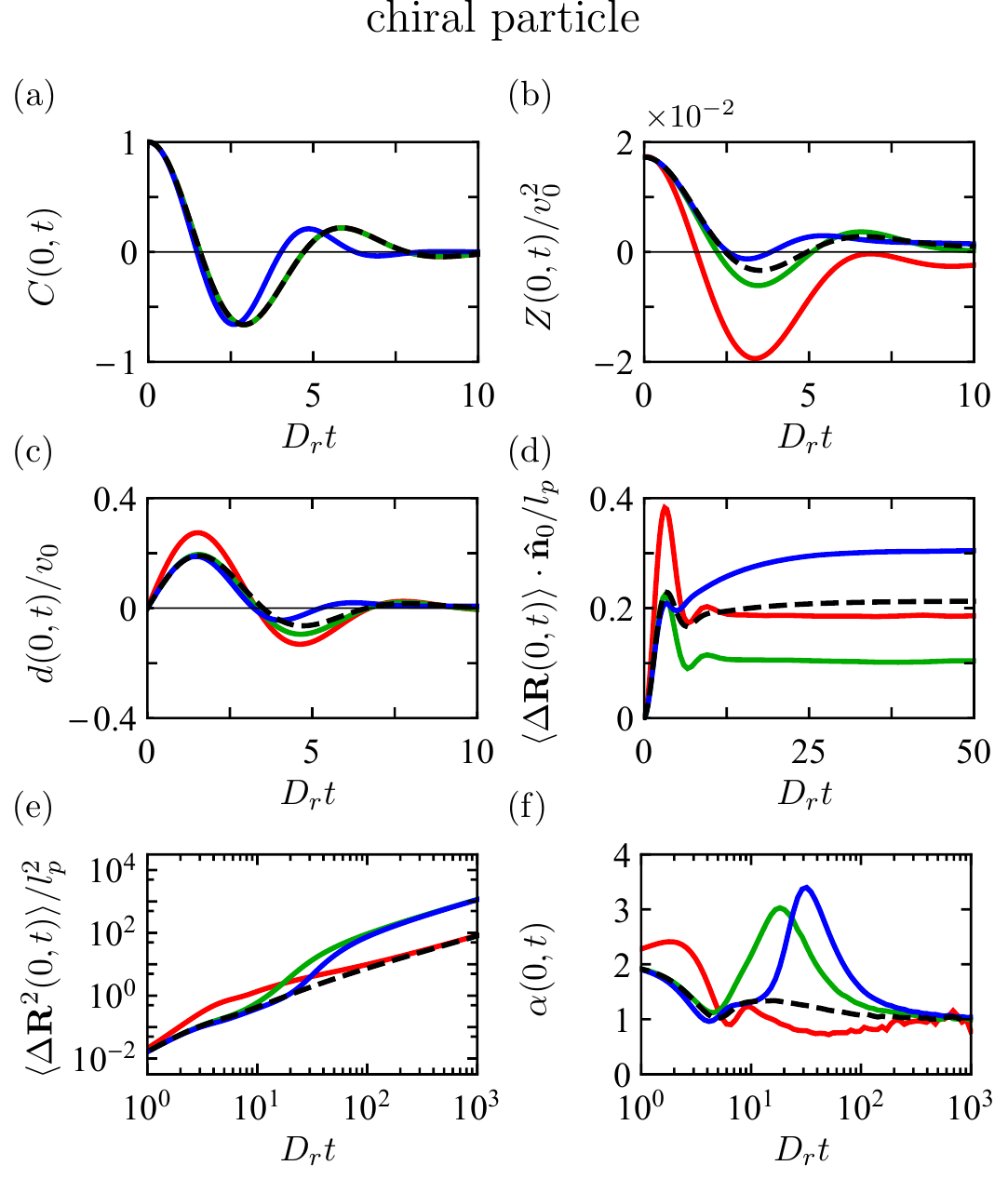}
	\caption{\label{fig:inertia_chiral}  
		Same as in Figure \ref{fig:inertia_achiral} for a chiral particle with a spinning frequency of $\omega=0.1 D_r$.}
\end{figure}

Numerical data for the special case of an exponential mass loss (see Eq.~\eqref{eq:m(t)_decay_exp}) and/or an exponential decay of the moment of inertia (see Eq.~\eqref{eq:J(t)_decay_exp}) as summarized in Table~\ref{table:parameter} are presented in Figures~\ref{fig:inertia_achiral} and \ref{fig:inertia_chiral}.
Figure~\ref{fig:inertia_achiral} is for an achiral particle and Figure~ \ref{fig:inertia_chiral} for a chiral particle. 
The case of time-independent inertia (with the parameters at time $t=0$) is shown as a reference, too. 
Equations \eqref{eq:general_model_trans} and \eqref{eq:general_model_rot} were discretized to perform Brownian dynamics simulations. 
For these simulations, we chose the time step $\Delta t = 10^{-2}/D_r$ and we performed $10^{6}$ realizations to calculate the respective ensemble averages.

We first discuss the case of an achiral particle. 
For isotropic shape change, the orientational correlation function $C(0,t)$ decorrelates faster (see Figure~\ref{fig:inertia_achiral}(a)), since the rotational noise is amplified during the decay of the moment of inertia.
The velocity autocorrelation $Z(0,t)$ as well as the delay function $d(0,t)$  decorrelate faster if the particle actually looses mass (see Figure~\ref{fig:inertia_achiral}(b) and \ref{fig:inertia_achiral}(c)).
For the particle with directed mass ejection, we see an increase in the velocity autocorrelation for short times and a more pronounced peak in the delay function due to the additional acceleration which enhances the particle velocity. 
The mean displacement along the initial displacement $\mean{\Delta \vec{R}(0,t)} \cdot \uvec{n}_0$ is displayed in Figure~\ref{fig:inertia_achiral}(d). 
Although the particle with directed mass ejection is the fastest for short times, it gets overtaken for long times by the particle with time-independent inertia.
Last we discuss the mean-square displacement $\mean{\Delta\vec{R}^2(0,t)}$. 
Beside additional acceleration for the particle with directed mass ejection for short times, the long time diffusivity is identical to the case of time-independent inertia.
In contrast, the cases of isotropic mass evaporation and isotropic shape end up with a decreased long-time diffusion coefficient (see Figure~\ref{fig:inertia_achiral}(e)) due to a smaller persistence.
The differences between the set-ups become clearer by considering the logarithmic derivative of the mean-square displacement
\begin{equation}\label{eq:alpha}
\alpha(t_1,t_2)  = \tdif{ \ln \big( \mean{\Delta \vec{R}^2(t_1,t_2)} \big) }{\ln(t_2)},
\end{equation}
If the mean-square displacement follows a power law $\mean{\Delta \vec{R}^2(t_1,t_2)} \sim (t_2-t_1)^\alpha $, $\alpha(t_1-t_2)$ is equal to the power-law exponent $\alpha$. 
This scaling exponent is shown in Figure~\ref{fig:inertia_achiral}(f).
All set-ups start in a ballistic regime ($\alpha =2$) for short times and end up in a diffusive regime ($\alpha = 1$) for long times.
Again for the particle with directed mass ejection we observe faster motion for short times indicated by a
super-ballistic scaling $\alpha > 2$ due to the acceleration. 
For times greater than the inverse decay rate of the moment of inertia $1/\gamma_{J}$, the particle with isotropic mass evaporation and an isotropic shape change behave sub-diffusively with $\alpha<1$  since their effective diffusivity decreases. 

Now we turn to the case of a chiral particle. 
First of all, even for constant parameters, the presence of the torque $M$ yields systematic oscillations in the orientation and velocity autocorrelations, and also in the delay function (see Figure~\ref{fig:inertia_chiral}(a)-\ref{fig:inertia_chiral}(c)). 
Indeed such oscillations in the delay function have been found recently in data for whirligig beetles \cite{Turner2019}. 
Turning to the time-dependent cases, similar to the pirouette of figure skating, the particle with an isotropic shape contraction is spinning with a higher frequency during the decay of the moment of inertia.
This is visible in the orientational and velocity autocorrelation functions and the delay function (see Figure~\ref{fig:inertia_chiral}(a)-\ref{fig:inertia_chiral}(c)).
Also, when the particle loses mass, the oscillation becomes more pronounced since the particle can adapt more easily to orientation changes. 
In contrast to the achiral case, the long-time behavior of the mean-square displacement increases for the time-dependent set-ups when the moment of inertia $J(t)$ decreases (see Figure~\ref{fig:inertia_chiral}(e)) in line with the trend discussed previously in Figure~\ref{Long_time_MSD_Fig}. 
This is marked by a peak in the scaling exponent for times larger than $1/\gamma_{J}$ (see Figure~\ref{fig:inertia_chiral}(f)).

\subsection{\label{sec:slowly_varying}Adiabatic approximation for slow variations}

When the parameters (such as mass $m(t)$, moment of inertia $J(t)$, friction coefficients $\xi(t)$ and $\xi_r(t)$, noise strengths $D(t)$ and $D_r(t)$, self-propulsion velocity $v_0(t)$) change very slowly in time, i.e.\ much slower than any other time-scale inherent in the model, the system can be analyzed using the adiabatic approximation.
In other words, one can take the expressions for the dynamical correlation function with constant parameters (as discussed in section \ref{sec:constant_parameters}) and insert into these expressions the slowly varying  time-dependent parameters.
This approximation becomes exact if the two time scales (largest system time scale and fastest time scale governing the change of all parameters) are separated completely.

Let us elaborate on the adiabatic approximation for the mean-square displacement (MSD) by considering an achiral active particle.
Corresponding analytical expressions for the MSD in the two limits of small and high moments of inertia $J$ are given by \eqref{eq:Long_time_diffusion_coefficient_smallJ} and \eqref{eq:Long_time_diffusion_coefficient_bigJ} respectively.
Using the long-time limit \eqref{eq:MSD_long_time_limit} we obtain within the adiabatic approximation for large $J$
\begin{equation} \label{eq:MSD_bigJ_time_dependent_parameters}
\mean{ \Delta\vec{R}^2(t) } \sim 4  \Bigg( D(t) +  \frac{v_0^{2}(t)}{4} \sqrt{\frac{2 \pi J(t)}{D_{r}(t)\,\xi_{r}(t)}}  \Bigg) t,
\end{equation}
when the moment of inertia $J$ becomes sufficiently large and
\begin{equation} \label{eq:MSD_smallJ_time_dependent_parameters}
\mean{  \Delta\vec{R}^2(t) }\sim 4 \Bigg( D(t) +\frac{v_0^{2}(t)}{2 D_{r}(t)} + \frac{v_0^{2}(t) J(t)}{2\xi_{r}(t)} \Bigg) t,
\end{equation}
in the case of a small moment of inertia $J$.
Let us now assume  a  slow power law in time for the moment of inertia, the self-propulsion, the rotational friction, and the diffusion coefficients
\begin{equation}\label{power_law_time_dependence}
v_0(t) \sim t^\beta,\quad J(t) \sim t^\delta,\quad \xi_r(t) \sim t^\varepsilon, \quad D_r(t) \sim t^\eta,
\end{equation}
with prescribed dynamical exponents $\beta$, $\delta$, $\epsilon$ and $\eta$. Plugging this into the expressions \eqref{eq:MSD_bigJ_time_dependent_parameters} and \eqref{eq:MSD_smallJ_time_dependent_parameters} we obtain
a power law for the  long-time MSD of the active particle
\begin{equation}
\mean{  \Delta\vec{R}^2(t) } \sim  t^{\alpha},
\label{MSD_bigJ_time_dependentJ}
\end{equation}
with
\begin{equation} \label{alpha}
\alpha=\max(1,1 + 2\beta - \frac{1}{2} \left( \varepsilon -\delta + \eta \right)),
\end{equation}
for large $J$ and
\begin{equation} \label{alpha_time_smallJ_dependentJ}
\alpha= \max(1,1 + 2 \beta - \min( \varepsilon -\delta , \eta)),
\end{equation}
for small $J$.
If $\alpha > 1$, adiabatic term  is dominating overwhelming the standard diffusion such that the particle  exhibits
{\it anomalous super-diffusion}.
If  $\alpha = 1$, the full MSD is dominated by the translational diffusion.
We finally remark that simpler scaling laws were obtained earlier in the overdamped limit \cite{Babel_2014}.

\section{\label{sec:conclusion} Conclusions}

To conclude, we have investigated the dynamics of an inertia-dominated Brownian particle, referred to as active Langevin dynamics.
Dynamical correlations within a simple model were calculated for a single "microflyer'' which is  simultaneously subjected to self-propulsion, inertia, damping and fluctuations and analytical results known for the overdamped limit of microswimmers were generalized to the inertial situation.
In particular, we considered the case of time-dependent inertia. 
Furthermore we identified a basic Langevin model for a rocket-like particle self-propelled by the ejection of mass for which we calculated its mean reach and found a noise-induced discontinuous transition in the optimal propulsion strategy for reaching the furthest distance.
The case of chiral particles referred to as circle-flyers was included.
One characteristic dynamical correlation absent in the overdamped case concerns the inertial delay between the orientation variations and the subsequent changes in the velocity direction.
For achiral particles with vanishing spinning frequency,  the inertial delay decays to zero after a profound peak at a typical delay time.
Conversely, for chiral particles, the inertial delay correlation may oscillate between positive and negative values.
Finally we have also addressed the limiting "adiabatic" case of very slow inertia  variation and have highlighted that a microflyer can undergo anomalous diffusion if the parameters are varying as a power law in time.

Future work should generalize the present model to external potentials like optical fields, disorder  and confinement \cite{BianchiPVMDL2016,CapriniMPV2019,DauchotD2019,MilitaruIFTND2020,BreoniSL2020} and to motion in noninertial rotating frames \cite{Loewen2019,ZhengL2020}.
Furthermore, anisotropic particles which show out-of-plane orientations and positions relevant for active complex plasmas \cite{LandauLifshitz1976} should be considered in the future.
In this case, the equations of motion are getting more complex involving friction and inertia tensors significantly more complicated than in the overdamped limit \cite{WittkowskiL2012,KraftWtHEPL2013}.
Next, the  "rocket-like" particles studied here should be realized in experiments, the most promising way seems to be dust particles in the plasma with evaporating mass.
Moreover it would be interesting to study collective effects of inertia-dominated active particles such as motility-induced-phase-separation  \cite{SumaGMO2014,ScholzRotor2018,PetrelliDCGS2018,MayyaNSHE2019,MandalLL2019,CapriniM2020b} or pattern formation in general \cite{AroldS2020}. Finally it would be interesting to generalize the more coarse-grained Ornstein-Uhlenbeck model for inertial active particles \cite{CapriniM2020a,NguyenWL2020} to the situation of time-dependent parameters.

\section{Acknowledgments}

We thank Christian Scholz, Ian Williams and Anton Ldov for helpful discussions. This work is supported by the German Research Foundation through grants LO 418/23-1 and IV 20/3-1.

\appendix

\section{\label{sec:general_solution}General solution}

For an analytical solution of the equations of motion we first consider the rotational part (see Eq.~\eqref{eq:general_model_rot})
For  $\phi_0 = \phi(t=0)$ and $\dot\phi_0 = \dot\phi(t=0)$, the solution of Eq.~\eqref{eq:general_model_rot} is
\allowdisplaybreaks{
\begin{align} \label{phidotIntegralForm}
\dot\phi(t) &  =  \dot{\phi}_{0} \, e^{-\Gamma_r(0,t)} \\
& + \integral{0}{t}{t'} \frac{\xi_r(t')}{J(t')} \, \omega(t') \,  e^{-\Gamma_r(t',t)}  \nonumber \\
& +  \integral{0}{t}{t'} \frac{\xi_r(t')}{J(t')} \sqrt{2D_{r}(t')} \tau_{\rm st}(t') \,  e^{-\Gamma_r(t',t)} , \nonumber 
\end{align}}
and thus
\begin{align} \label{phiIntegralForm}
\phi(t) &=  \phi_{0} + \integral{0}{t}{t'} \dot{\phi}_{0} \, e^{-\Gamma_r(0,t')} \\
& +  \integral{0}{t}{t'} \integral{0}{t'}{t''} \frac{\xi_r(t'')}{J(t'')} \,  \omega(t'')   e^{-\Gamma_r(t'',t')} \nonumber \\
& +  \integral{0}{t}{t'} \integral{0}{t'}{t''} \frac{\xi_r(t'')}{J(t'')} \sqrt{2D_{r}(t'')} \tau_{\rm st}(t'')  e^{-\Gamma_r(t'',t')}, \nonumber
\end{align}
where
\begin{equation}
\Gamma_r(t_1,t_2) = \int_{t_1}^{t_2} dt' \, \frac{\xi_r(t')}{J(t')} + (1- \nu) \ln \left( \frac{J(t_2)}{J(t_1)} \right).
\end{equation}

The translational equation of motion yields for the particle velocity  
\begin{align} \label{velx_lab}
\dot{\vec{R}}(t) &  =  \dot{\vec{R}}_{0} \, e^{-\Gamma(0,t)} \\
& + \integral{0}{t}{t'} a(t') \, \uvec{n}(t')  \,  e^{-\Gamma(t',t)}  \nonumber \\
& +  \integral{0}{t}{t'}  \frac{\xi(t')}{m(t')} \sqrt{2D(t')} \vec{f}_{\rm st}(t')  \,  e^{-\Gamma(t',t)}. \nonumber 
\end{align}
Hence, the center-of-mass position is calculated as
\begin{align}\label{x_lab}
\vec{R}(t) &=  \vec{R}_{0} + \integral{0}{t}{t'} \dot{\vec{R}}_{0} \, e^{-\Gamma(0,t')} \\
& +  \integral{0}{t}{t'} \integral{0}{t'}{t''} a(t'') \,\uvec{n}(t'')  e^{-\Gamma(t'',t')} \nonumber \\
& +  \integral{0}{t}{t'} \integral{0}{t'}{t''}  \frac{\xi(t'')}{m(t'')} \sqrt{2D(t'')} \vec{f}_{\rm st}(t'')   e^{-\Gamma(t'',t')}, \nonumber
\end{align}
where
\begin{equation}
a(t) = \frac{\xi(t)}{m(t)} v_{0}(t) - \frac{\dot{m}(t)}{m(t)} u(t).
\end{equation}
and
\begin{equation}
\Gamma(t_1,t_2) = \int_{t_1}^{t_2} dt' \, \frac{\xi(t')}{m(t')}. 
\end{equation}
Here $\vec{R}_{0}$ and $\dot{\vec{R}}_{0}$ are the initial position and velocity at time $t=0$.

\section{\label{sec:C}Exponential mass loss}

In an isothermal environment of temperature $T$, the mass loss through a small leak of cross section $S$ in the rocket tank of volume $V$ in quasi-equilibrium is governed by
\begin{equation}
\dot{m}_\text{fuel}(t) = - \frac{1}{6} \frac{S}{V}\sqrt{\frac{3 k_B T}{ m_\text{mol} }} \, m_\text{fuel}(t)=-\gamma_m \, m_\text{fuel}(t),
\label{m(t)_exp}
\end{equation}
where $ m_\text{mol}$ is the mass of the ejected molecules and $k_B$ is the Boltzmann constant. 
Equation~\eqref{m(t)_exp} implies an exponential decay of the rocket fuel, i.e., $ m_\text{fuel}(t) = m_\text{fuel}(0) \,  e^{-\gamma_m t}$ with the mass decay rate $\gamma_{m}$
and thus motivates Eq.~\eqref{eq:m(t)_decay_exp}.

\bibliographystyle{apsrev4-1}
\bibliography{refs}

\end{document}